\documentclass[10pt,prd,singlecolumn,
nofootinbib,
noshowkeys,
noshowpacs,
superscriptaddress,
floatfix
]{revtex4-2}
\usepackage{amsmath,amsfonts,amsthm,amssymb}
\usepackage[dvips]{graphics,graphicx}
\usepackage[usenames,dvipsnames]{color}
\definecolor{darkblue}{RGB}{0,0,196}
\definecolor{darkgreen}{RGB}{0,120,0}
\usepackage[colorlinks=true,linktocpage=true,linkcolor=darkblue,citecolor=red,urlcolor=darkblue]{hyperref}
\usepackage{cancel}
\usepackage{bbold}
\usepackage{multirow}
\usepackage{longtable}
\usepackage{color}
\usepackage[normalem]{ulem}
\usepackage{hyperref}
\usepackage{bigints}
\usepackage{xparse}
\usepackage{physics}
\usepackage{verbatim}
\usepackage{minibox}
\usepackage{comment}
\usepackage{appendix}
\usepackage{slashed}
\usepackage{marginnote}
\usepackage{graphicx}
\usepackage[nice]{nicefrac}
\usepackage{amsmath}
\usepackage[colorinlistoftodos]{todonotes}
\usepackage{hepunits}
\usepackage{stackengine,scalerel}
\newcommand\hstar[1]{\ThisStyle{\ensurestackMath{%
  \setbox0=\hbox{$\SavedStyle#1$}%
  \stackengine{0pt}{\copy0}{\kern.2\ht0\smash{\SavedStyle\star}}{O}{c}{F}{T}{S}}}}

\definecolor {darkgreen}{rgb}{0.2,0.7,0.2}

\begin{document}

\title{Kubo formula for spin hydrodynamics: spin chemical potential as leading order in gradient expansion}

\author{Sourav Dey \footnote{sourav.dey@niser.ac.in}}

\address{School of Physical Sciences, National Institute of Science Education and Research, An OCC of
Homi Bhabha National Institute, Jatni-752050, India}

\author{Arpan Das \footnote{arpan.das@pilani.bits-pilani.ac.in}}

\address{Department of Physics,  Birla Institute of Technology and Science Pilani, Pilani Campus, Pilani,  Rajasthan-333031, India}


\begin{abstract}
We present a first-order dissipative spin hydrodynamic framework, where the spin chemical potential $\omega^{\mu\nu}$ is treated as the leading term in the hydrodynamic gradient expansion, i.e., $\omega^{\mu\nu}\sim \mathcal{O}(1)$. We argue that for the consistency of the theoretical framework, the energy-momentum tensor needs to be symmetric at least up to order $\mathcal{O}(\partial)$. We consider the phenomenological form of the spin tensor, where it is anti-symmetric in the last two indices only. A comprehensive analysis of spin hydrodynamics is conducted using both macroscopic entropy current analysis and microscopic Kubo formalism, establishing consistency between the two approaches. A key finding is the entropy production resulting from spin-orbit coupling, which alters the traditional equivalence between the Landau and Eckart fluid frames. Additionally, we identify cross-diffusion effects, where vector dissipative currents are influenced by gradients of both spin chemical potential and chemical potential corresponding to the conserved charge through off-diagonal transport coefficients. Two distinct methods for decomposing the spin tensor are proposed, and their equivalence is demonstrated through Kubo relations.
\end{abstract}

\maketitle
\section{Introduction}
The observation of spin polarization of hadrons in non-central relativistic heavy-ion collisions has marked the onset of QCD-spintronics, i.e., spin dynamics in an evolving QCD medium.~\cite{STAR:2017ckg, STAR:2018gyt,STAR:2019erd,ALICE:2019onw,ALICE:2019aid,STAR:2020xbm,Kornas:2020qzi,STAR:2021beb,ALICE:2021pzu}. In the non-central heavy ion collision, after the collision event, it is expected that most of the initial angular momentum of the colliding nuclei will be carried away by the spectators. However, a sizable fraction of the initial angular momentum can remain in the created thermalized partonic medium due to the inhomogeneous density profile of the colliding nuclei. Such a nonvanishing angular momentum can potentially give rise to nonzero vorticity or rotational structure in the QCD medium~\cite{Jiang:2016woz}. One of the novel features of the presence of fluid vorticity is the spin-vorticity coupling that can polarize the particles with non-zero spin~\cite {Liang:2004ph,Becattini:2007sr,Gao:2007bc,Huang:2011ru,Becattini:2013fla,Fang:2016vpj,Voloshin:2004ha,Betz:2007kg}. Spin polarization of hadrons has been indeed observed in the heavy ion collision experiments across a wide range of center of mass energy~\cite{STAR:2017ckg,ALICE:2019onw,Kornas:2020qzi}.  Experimentally there are two distinct spin observable. The first one is the measure of spin polarization along the direction of global angular momentum (direction normal to the reaction plane of heavy ions). This observation is termed as the  {\it global spin polarization}. The second spin observable is the measurement of the longitudinal spin polarization, i.e., variation of the spin polarization along the beam direction with the azimuthal angle in the transverse plane~\cite{Becattini:2020ngo}. The longitudinal spin polarization observation is also known as the {\it local spin polarization}. 
Various theoretical models including different relativistic dissipative hydrodynamic models~\cite{DelZanna:2013eua,Karpenko:2013wva,Ivanov:2019ern}, parton cascade model (AMPT)~\cite{Li:2017slc}, hadronic cascade model (UrQMD)~\cite{Vitiuk:2019rfv}, chiral kinetic theory~\cite{Sun:2017xhx}, etc. have been used to understand the global and local spin polarization observation. All these different models take into account the spin-thermal vorticity coupling and have not been able to provide a satisfactory unified explanation of all the polarization measurements without invoking additional assumptions. In particular, models that incorporate only the spin-vorticity coupling predict an opposite azimuthal angle dependence of the longitudinal spin polarization as compared to the experimental observation. This is usually called the ``spin sign problem"~\cite{Becattini:2020ngo}. To resolve this problem, new approaches have been developed, e.g., thermal shear-induced corrections~\cite{Liu:2020dxg,Yi:2021ryh,Sun:2021nsg,Florkowski:2021xvy},
chiral kinetic theory (CKT) based model~\cite{Hidaka:2017auj}, etc. which have provided a new perspective to the longitudinal spin polarization problem. For interesting discussions on spin polarization in heavy ion collision, please see Refs.~\cite{Becattini:2020ngo,Becattini:2022zvf,Gao:2020vbh, Becattini:2020sww,Wang:2017jpl}. One of the approaches that triggered a lot of investigation is the spin-hydrodynamic framework for the dynamical explanation of the spin sign problem~\cite{Florkowski:2018ahw,Florkowski:2017dyn,Florkowski:2017ruc,Florkowski:2019qdp,Florkowski:2019voj,Hattori:2019lfp,Fukushima:2020ucl,Li:2020eon,She:2021lhe,Montenegro:2017lvf,Montenegro:2017rbu,Florkowski:2017ruc,Florkowski:2018myy,Bhadury:2020puc,Shi:2020qrx,Weickgenannt:2022zxs,Weickgenannt:2019dks,Weickgenannt:2020aaf,Speranza:2021bxf,Wang:2021ngp,Gallegos:2021bzp,Wang:2021ngp,Gallegos:2021bzp}.  This apart, relativistic kinetic theory approach~\cite{Florkowski:2017ruc,Florkowski:2017dyn,Hidaka:2017auj,Florkowski:2018myy,Weickgenannt:2019dks,Bhadury:2020puc,Weickgenannt:2020aaf,Shi:2020htn,Speranza:2020ilk,Bhadury:2020cop,Singh:2020rht,Bhadury:2021oat,Peng:2021ago,Sheng:2021kfc,Sheng:2022ssd,Hu:2021pwh,Hu:2022lpi,Fang:2022ttm,Wang:2022yli,Bhadury:2024ckc}, effective Lagrangian approach~\cite{Montenegro:2017rbu,Montenegro:2017lvf,Montenegro:2018bcf,Montenegro:2020paq}, quantum statistical density operator approach~\cite{Becattini:2007nd,Becattini:2009wh,Becattini:2012pp,Becattini:2012tc,Becattini:2018duy,Hu:2021lnx}, holographic methods~\cite{Gallegos:2020otk,Garbiso:2020puw}, etc. have been considered to study the spin polarization in a vortical QCD plasma. 

In this article, we consider a first-order dissipative spin hydrodynamic framework where the spin chemical potential is the leading order in the expansion of the hydrodynamic gradient $\omega^{\mu\nu}\sim \mathcal{O}(1)$.
In the literature, spin hydrodynamic framework considering both $\omega^{\mu\nu}\sim \mathcal{O}(1)$ and $\omega^{\mu\nu}\sim \mathcal{O}(\partial)$, have been developed~\cite{Hattori:2019lfp,Biswas:2023qsw,Biswas:2022bht,Daher:2022xon,Daher:2022wzf,She:2021lhe}.  It can be argued that if the energy-momentum tensor has an anti-symmetric part, then spin chemical potential $\omega^{\mu\nu}$ is completely determined in terms of thermal vorticity  $\varpi^{\mu\nu}\equiv -(\partial_{\mu}(u_{\nu}/T)-\partial_{\nu}(u_{\mu}/T))/2$ in global equilibrium~\cite{Becattini:2013fla,Florkowski:2018fap,Rindori:2020qqa}. However, in local equilibrium, such a requirement is not necessary. Hence in spin hydrodynamic frameworks, where one assumes the local thermodynamic equilibrium $\omega^{\mu\nu}$ can be considered as an independent field having its own evolution consistent with the conservation equations. If the energy-momentum tensor is asymmetric, then one expects that in global equilibrium, the spin chemical potential and the thermal vorticity will be related. Since thermal vorticity is defined as, $\varpi^{\mu\nu}\equiv -(\partial_{\mu}(u_{\nu}/T)-\partial_{\nu}(u_{\mu}/T))/2 \sim \mathcal{O}(\partial)$ it is natural to consider that $\omega^{\mu\nu}\sim \mathcal{O}(\partial)$~\cite{Hattori:2019lfp}. Using this hydrodynamic gradient scheme first-order~\cite{Hattori:2019ahi,Fukushima:2020ucl,Daher:2022wzf,Daher:2022xon,Biswas:2022bht} and second-order spin hydrodynamic framework~\cite{Biswas:2023qsw} has been developed. However, the hydrodynamic gradient ordering of $\omega^{\mu\nu}$ is not unique, particularly when the energy-momentum tensor is symmetric. If the energy-momentum tensor is symmetric, then even in the global equilibrium, we can not express the spin chemical potential in terms of the thermal vorticity. In such a situation, we can consider the spin chemical potential in the leading order in the hydrodynamic gradient expansion, i.e., $\omega^{\mu\nu}\sim\mathcal{O}(1)$. $\omega^{\mu\nu}\sim\mathcal{O}(\partial)$, and $\omega^{\mu\nu}\sim\mathcal{O}(1)$ 
are two entirely different situations, giving rise to different spin hydrodynamic frameworks~\cite{Hattori:2019lfp,Biswas:2023qsw,Biswas:2022bht,Daher:2022xon,Daher:2022wzf}. 

In this article, by considering the spin chemical potential as the leading order in the hydrodynamic gradient expansion, i.e., $\omega^{\alpha\beta}\sim \mathcal{O}(1)$, we first look into the first-order dissipative spin hydrodynamic framework. We show that the spin-orbit coupling gives rise to cross conductivity between charge current and heat (a net momentum flow) current. This is a novel feature of the present calculation. Moreover, applying Zubarev's non-equilibrium statistical operator method, we also obtain Kubo formulae for various transport coefficients, including the cross-conductivity coefficients that appear in the calculation. The non-equilibrium statistical operator method has been routinely used to obtain transport coefficients in relativistic hydrodynamic theories~\cite{Policastro:2001yc,Arnold:2000dr,Arnold:2003zc,Jeon:1994if,Jeon:1995zm,Wang:2024afv, Jaiswal:2024urq, Wagner:2024fry}, relativistic magnetohydrodynamic theories~\cite{Huang:2011dc}. Zubarev's approach has also been used to obtain Kubo relations for spin hydrodynamic theory where the author considered $\omega^{\alpha\beta}\sim \mathcal{O}(\partial)$~\cite{Hu:2021lnx,Tiwari:2024trl}. In the non-equilibrium statistical operator method, one assumes that the system is not far away from the equilibrium so that we can apply the gradient expansion and the linear response theory to calculate the Green-Kubo relation for different transport coefficients. 

In this manuscript, we have used the following notations and conventions. We use the metric tensor with mostly negative signature $g_{\mu\nu}= \hbox{diag}(+1, -1, -1, -1)$ and the totally antisymmetric Levi-Civita tensor with the sign convention $\epsilon^{0123} = -\epsilon_{0123} = 1$. $ u^{\mu}$ represents the fluid four-velocity, which is normalized to unity, i.e., $u^{\mu}u_{\mu}= 1$. The symmetric and antisymmetric combinations are represented as $A^{\{\alpha}B^{\beta\}}=(A^{\alpha}B^{\beta}+A^{\beta}B^{\alpha})/2$ and $A^{[\alpha}B^{\beta]}=(A^{\alpha}B^{\beta}-A^{\beta}B^{\alpha})/2$, respectively.  The projector  $\Delta^{\mu\nu}\equiv g^{\mu\nu}-u^{\mu}u^{\nu}$ is orthogonal to $u^{\mu}$, i.e., $\Delta^{\mu\nu}u_{\mu}=0$. Projection of a four vector $V^{\mu}$ orthogonal to $u^{\mu}$ is represented as $V^{\langle\mu\rangle}\equiv \Delta^{\mu\nu}V_{\nu}$. Traceless and symmetric projection operator orthogonal to flow vector is denoted as $A^{\langle\mu}B^{\nu\rangle}\equiv \Delta^{\mu\nu}_{\alpha\beta}A^{\alpha}B^{\beta}\equiv \frac{1}{2}\left(\Delta^{\mu}_{~\alpha}\Delta^{\nu}_{~\beta}+\Delta^{\mu}_{~\beta}\Delta^{\nu}_{~\alpha}-\frac{2}{3}\Delta^{\mu\nu}\Delta_{\alpha\beta}\right)A^{\alpha}B^{\beta}$. Similarly, $A^{\langle[\mu}B^{\nu]\rangle}\equiv \Delta^{[\mu\nu]}_{[\alpha\beta]}A^{\alpha}B^{\beta}\equiv \frac{1}{2}\left(\Delta^{\mu}_{~\alpha}\Delta^{\nu}_{~\beta}-\Delta^{\mu}_{~\beta}\Delta^{\nu}_{~\alpha}\right)A^{\alpha}B^{\beta}$ represents the antisymmetric projection operator orthogonal to the flow vector. The partial derivative operator ($\partial_{\mu}$) can be decomposed into one part which is along the flow direction ($D\equiv u^{\mu}\partial_{\mu}$) and another part which is orthogonal to the flow direction ($\nabla_{\mu}\equiv\Delta_{\mu}^{~\alpha}\partial_{\alpha}$), i.e., $\partial_{\mu}=u_{\mu}D+\nabla_{\mu}$. Here, $D$  denotes the comoving derivative. The fluid expansion rate is defined as $\theta\equiv \partial_{\mu}u^{\mu}=\nabla_{\mu}u^{\mu}$. The symmetric traceless combination of the derivative of the fluid flow is defined as $\sigma_{\mu\nu}\equiv\frac{1}{2}(\nabla_{\mu} u_{\nu}+\nabla_{\nu} u_{\mu})-\frac{1}{3}\theta\Delta_{\mu\nu}$.

The rest of the paper is arranged in the following way. In Sec.~\ref{sec1}, we outline the spin hydrodynamic framework through an entropy current analysis. Subsection~\ref{sec1a} introduces the macroscopic conservation equations relevant to spin hydrodynamics. In subsection~\ref{entropycurrentapp}, we provide the tensor decomposition of the various dissipative currents that emerge in the framework. Using entropy current analysis, we derive the constitutive relations for the dissipative currents and identify the associated transport coefficients. In Sec.~\ref{sec2c}, we propose an alternative method for decomposing the dissipative part of the spin tensor, which introduces additional transport coefficients. We also discuss the equivalence between different decomposition schemes. In Sec.~\ref{sec3}, we examine the nonequilibrium statistical operator method to derive the Kubo relations for the transport coefficients in spin hydrodynamics. This section also demonstrates the equivalence of different tensor decomposition methods by explicitly showing the mathematical connections between the transport coefficients appearing in these approaches. Finally, in Sec.~\ref{sec4}, we conclude with a summary of our findings and offer an outlook on future works.

\section{Spin hydrodynamic framework: entropy current analysis}
\label{sec1}
\subsection{Macroscopic conservation equations}
\label{sec1a}
Spin hydrodynamic frameworks are based on the macroscopic conservation of $T^{\mu\nu}$, $J^{\mu}$, and $J^{\mu\alpha\beta}$, 
\begin{align}
& \partial_{\mu}T^{\mu\nu}=0, ~~~~\partial_{\mu}J^{\mu}=0,~~~\partial_{\lambda}J^{\lambda\mu\nu}=0,\label{equ1ver1}
\end{align}
here  $T^{\mu\nu}$ is the energy-momentum tensor, $J^{\mu}$ is the global conserved current, e.g., net baryon number current in QCD, etc., and $J^{\mu\alpha\beta}$ is the total angular momentum tensor~\cite{Florkowski:2018fap}. The total angular momentum tensor $J^{\lambda\mu\nu}$ can be expressed in terms of the orbital part ($L^{\lambda\mu\nu}$) and spin part ($S^{\lambda\mu\nu}$), 
\begin{align}
&J^{\lambda\alpha\beta}=L^{\lambda\alpha\beta}+S^{\lambda\alpha\beta}; ~~~L^{\lambda\alpha\beta}=x^{\alpha}T^{\lambda\beta}-x^{\beta}T^{\lambda\alpha}.\label{equ2ver1}
\end{align}
$J^{\lambda\alpha\beta}$, $L^{\lambda\alpha\beta}$, and $S^{\lambda\alpha\beta}$ are antisymmetric in last two indices. In general neither $L^{\lambda\mu\nu}$ nor $S^{\lambda\mu\nu}$ are separately conserved. The  conservation of the total angular momentum tensor implies the conservation or the non-conservation of the spin tensor, 
\begin{align}
\partial_{\lambda}S^{\lambda\mu\nu}=-T^{\mu\nu}+T^{\nu\mu}=-2 T^{[\mu\nu]}.\label{equ3ver1}     
\end{align}
It is evident from the above equation that for a symmetric energy-momentum tensor, the spin tensor is separately conserved. The anti-symmetric part gives rise to the spin-orbit conversion, which spoils the conservation of the spin tensor. We emphasize that the physical measurable quantities, $T^{\mu\nu}$, $J^{\mu}$, $S^{\lambda\mu\nu}$ are expectation values of the underling QFT operators $\widehat{T}^{\mu\nu}$, $\widehat{J}^{\mu}$, $\widehat{S}^{\lambda\mu\nu}$ respectively~\cite{Kovtun:2019hdm}. In general $T^{\mu\nu}$, $J^{\mu}$, and $S^{\lambda\mu\nu}$ contains dissipative terms. In standard hydrodynamics, if we ignore the dissipative corrections, then $T^{\mu\nu}$ and $J^{\mu}$ can be completely specified by temperature ($T$), chemical potential ($\mu$), and fluid flow ($u^{\mu}$). Together $T$, $\mu$, and $u^{\mu}$ have five degrees of freedom. Dynamics of $T$, $\mu$, and $u^{\mu}$ is determined by the conservation of $T^{\mu\nu}$ and $J^{\mu}$, i.e., five dynamical equations. In spin hydrodynamics, we have an additional six equations stemming from the conservation of $J^{\lambda\mu\nu}$. These six equations determine the dynamics of another anti-symmetric tensor having six degrees of freedom. This anti-symmetric tensor can be identified as the spin chemical potential ($\omega^{\mu\nu}$) analogous to the chemical potential ($\mu$) associated with the conserved current ($J^{\mu}$)~\cite{Florkowski:2018fap}\footnote{We emphasize that unlike the chemical potential ($\mu$) associated with the conserved current, the spin chemical potential does not imply the conservation of the spin tensor. Rather it can be considered as a Lagrange multiplayer that appears in the definition of the statistical density operator of the system with angular momentum~\cite{Florkowski:2018fap}.}. 

In this article, we do not consider $\omega^{\mu\nu}\sim \mathcal{O}(\partial)$, rather we consider $\omega^{\mu\nu}\sim \mathcal{O}(1)$. This is because if the energy-momentum tensor has an anti-symmetric part, then spin chemical potential $\omega^{\mu\nu}$ is completely determined in terms of thermal vorticity  $\varpi^{\mu\nu}\equiv -(\partial_{\mu}(u_{\nu}/T)-\partial_{\nu}(u_{\mu}/T))/2$ in global equilibrium~\cite{Becattini:2013fla,Florkowski:2018fap,Rindori:2020qqa}. This justifies the hydrodynamic gradient ordering of spin chemical potential as $\omega^{\mu\nu}\sim \mathcal{O}(\partial)$ ~\cite{Hattori:2019lfp,Biswas:2023qsw,Biswas:2022bht,Daher:2022xon,Daher:2022wzf}.  But if the energy-momentum tensor is symmetric, then  $\omega^{\mu\nu}\sim \mathcal{O}(\partial)$ is not obvious~\cite{She:2021lhe}. If the energy-momentum tensor is symmetric, then even in the global equilibrium, we can not express the spin chemical potential in terms of the thermal vorticity. In that case we can consider $\omega^{\mu\nu}\sim \mathcal{O}(1)$ to develop an alternative spin hydrodynamic framework~\cite{She:2021lhe,Daher:2022wzf}. We emphasize that the on-shell Wigner function solution (at zeroth order in $\hbar$ in the semi-classical expansion) for spin-half particles shows that the energy-momentum tensor remains symmetric in equilibrium~\cite{Das:2022azr, Speranza:2020ilk}. It is the off-shell corrections that drive the system out of equilibrium through microscopic collisions and can introduce an anti-symmetric component to the energy-momentum tensor~\cite{Weickgenannt:2021cuo}. This observation encourages us to explore a spin hydrodynamic framework where $\omega^{\mu\nu}\sim \mathcal{O}(1)$.

\subsection{Entropy current analysis: constitutive relation for dissipative currents}
\label{entropycurrentapp}

One of the conceptual difficulties that arises in frameworks of spin-hydrodynamics is the pseudo-gauge dependence. This implies that the canonical form of the energy-momentum tensor and spin tensor obtained using the Noether theorem are not unique~\cite{Chen:2018cts}. Given a pair of the energy-momentum tensor and the spin tensor, we can always define another set of these tensors using a pseudo-gauge transformation without affecting the conservation equations (Eq.~\eqref{equ1ver1})~\cite{Becattini:2013fla,Florkowski:2018fap}. 
Although the conservation equations are unaffected by the pseudo-gauge transformation the individual hydrodynamic currents are not pseudo-gauge transformation invariant which affects the spin hydrodynamic frameworks, e.g., canonical framework (Can)~\cite{Shi:2020qrx}, de Groot-van Leeuwen-van Weert framework (GLW)~\cite{DeGroot:1980dk}, etc. The pseudo-gauge dependence of the spin hydrodynamic frameworks is an intriguing problem that is still under intense investigation. For our purpose, we will consider the \textit{phenomenological} framework where $T^{\mu\nu}$ may consist of symmetric as well as antisymmetric parts \footnote{We are starting with an asymmetric energy-momentum tensor, and later we argue that the energy-momentum tensor needs to be symmetric when the spin chemical potential is $\mathcal{O}(1)$ in the hydrodynamic gradient expansion.} and the spin tensor has a simple \textit{phenomenological} form, which is only antisymmetric in the last two indices~\cite{Weyssenhoff:1947iua,Florkowski:2018fap,Florkowski:2017ruc}. Moreover, this framework can be argued to be thermodynamically consistent~\cite{Hattori:2019lfp}, and it can be obtained from the properly defined canonical tensors using pseudo-gauge transformation~\cite{Daher:2022xon}. In this framework, we start with the following constitutive relations for the energy-momentum tensor and the spin tensor~\cite{Hattori:2019lfp,Biswas:2023qsw,Biswas:2022bht,Daher:2022xon,Daher:2022wzf},
\begin{align}
    & T^{\mu\nu}_{} = T^{\mu\nu}_{(0)}+T^{\mu\nu}_{(1)}, ~~T^{\mu\nu}_{(0)}=\varepsilon u^{\mu}u^{\nu}-P\Delta^{\mu\nu}, \label{equ6ver1}\\
    & J^{\mu}_{} = J^{\mu}_{(0)}+J^{\mu}_{(1)}, ~~J^{\mu}_{(0)}=nu^{\mu},\label{equ7ver1}\\
    & S^{\mu\alpha\beta}_{}=S^{\mu\alpha\beta}_{(0)}+S^{\mu\alpha\beta}_{(1)},~~S^{\mu\alpha\beta}_{(0)}=u^{\mu}S^{\alpha\beta}.  \label{equ8ver1}
\end{align}  
Here $T^{\mu\nu}_{(0)}$, $J^{\mu}_{(0)}$, and $S^{\mu\alpha\beta}_{(0)}$ represent currents which are $\mathcal{O}(1)$ in the hydrodynamic gradient expansion. $T^{\mu\nu}_{(1)}$, $J^{\mu}_{(1)}$, and $S^{\mu\alpha\beta}_{(1)}$ represent $\mathcal{O}(\partial)$ terms, or the first order derivative corrections.  $T^{\mu\nu}_{(1)}$, $J^{\mu}_{(1)}$, and $S^{\mu\alpha\beta}_{(1)}$ satisfy the conditions $T^{\mu\nu}_{(1)}u_{\mu}u_{\nu}=0$, $J^{\mu}_{(1)}u_{\mu}=0$, and $S^{\mu\alpha\beta}_{(1)}u_{\mu}=0$ respectively.  $\varepsilon$ and $P$ are the energy density and pressure, respectively. The spin density tensor $S^{\mu\nu}$ is antisymmetric, i.e. $S^{\mu\nu}=-S^{\nu\mu}$. The tensor $S^{\alpha\beta}$ plays a role analogous to the number density ($n$)~\cite{Hattori:2019lfp,Fukushima:2020ucl}. Energy density ($\varepsilon$), pressure ($P$), number density ($n$), and spin density ($S^{\mu\nu}$) are all $\mathcal{O}(1)$ terms that can be written in terms of $\mathcal{O}(1)$ hydrodynamic variables, i.e, temperature ($T$), chemical potential ($\mu$), and spin chemical potential ($\omega^{\alpha\beta}$). The generalized laws of local thermodynamics, including the spin contribution, are~\cite{Hattori:2019lfp,Fukushima:2020ucl},
\begin{align} 
& \varepsilon+P =Ts+\mu n+\omega_{\alpha\beta}S^{\alpha\beta},\label{equ9ver1}\\
& d\varepsilon =Tds+\mu dn+\omega_{\alpha\beta}dS^{\alpha\beta},\label{equ10ver1}\\ 
& dP=sdT+n d\mu+S^{\alpha\beta}d\omega_{\alpha\beta}.
\label{equ11ver1}
\end{align}
Here $s (T,\mu,\omega^{\alpha\beta})$ is the entropy density, and the spin chemical potential  $\omega^{\mu\nu}$ is conjugate to the spin density $S^{\mu\nu}$~\cite{Hattori:2019lfp}. The above thermodynamic relation allowed us to write the ansatz for the covariant form of the entropy current, 
\begin{align}
\mathcal{S}^{\mu}=T^{\mu\nu}\beta_{\nu}+P\beta^{\mu}-\alpha J^{\mu}-\beta\omega_{\alpha\beta}S^{\mu\alpha\beta}.
\label{equ12ver1}
\end{align}
Here $\beta^{\mu}=\beta u^{\mu}=u^{\mu}/T$, and $\alpha=\mu/T$. The leading order term ($\mathcal{O}(1)$) in $\mathcal{S}^{\mu}$ is, 
\begin{align}
\mathcal{S}^{\mu}_{(0)} & =T^{\mu\nu}_{(0)}\beta_{\nu}+P\beta^{\mu}-\alpha J^{\mu}_{(0)}-\beta\omega_{\alpha\beta}S^{\mu\alpha\beta}_{(0)}\nonumber\\
& = \left(\varepsilon+P-\mu n-\omega_{\alpha\beta}S^{\alpha\beta}\right)\beta^{\mu}=su^{\mu}.
\label{equ13ver1}
\end{align}
The above relation justifies the use of Eq.~\eqref{equ12ver1} as the general form of the covariant entropy current, including the high-order derivative corrections. Using the spin hydrodynamic equations of motions, i.e., $\partial_{\mu}T^{\mu\nu}=0$, $\partial_{\mu}J^{\mu}=0$, and $\partial_{\mu}J^{\mu\alpha\beta}=0$ it can be shown that (see Appendix~\ref{appenA}),
\begin{align}
\partial_{\mu}\mathcal{S}^{\mu}_{(0)}=2\beta\omega_{\alpha\beta}T^{[\alpha\beta]}_{(1)}
\label{equ14ver1}
\end{align}
This is a very interesting result, and a few comments are in order here. In the standard hydrodynamics one expects that $\partial_{\mu}\mathcal{S}^{\mu}_{(0)}=0$, i.e., ideal fluid does not produces entropy. However, in spin hydrodynamics, the situation is a bit different. Note that in spin hydrodynamics, $\omega^{\alpha\beta}$ is non-zero. Therefore, any nonvanishing $T^{[\alpha\beta]}_{(1)}$ will spoil the existence of the equilibrium entropy current, i.e., $\partial_{\mu}\mathcal{S}^{\mu}_{(0)}\neq 0$. In the absence of spin chemical potential, this dilemma never appears. But if we demand that $\mathcal{S}^{\mu}_{(0)}$ corresponds to reversible ideal spin hydrodynamics, then $T^{[\alpha\beta]}_{(1)}$ must vanish, implying that anti-symmetric part in the energy-momentum tensor should only arise at second order or higher in gradient expansion~\cite{She:2021lhe}. The same conclusion can be achieved by looking into the divergence of $\mathcal{S}^{\mu}$ (for a detailed derivation, see Appendix~\ref{appenB}), 
\begin{align}
\partial_{\mu}\mathcal{S}^{\mu} & = (\partial_{\mu}\beta_{\nu})T^{\mu\nu}+\beta_{\nu}\partial_{\mu}T^{\mu\nu}+\beta^{\mu}\partial_{\mu}P+P\partial_{\mu}\beta^{\mu}\nonumber\\
& ~~~~~-J^{\mu}\partial_{\mu}\alpha-\alpha \partial_{\mu}J^{\mu}-S^{\mu\alpha\beta}\partial_{\mu}\Omega_{\alpha\beta}-\Omega_{\alpha\beta} \partial_{\mu}S^{\mu\alpha\beta}\nonumber\\
& = T^{\{\mu\nu\}}_{(1)}\partial_{\{\mu}\beta_{\nu\}}+T^{[\mu\nu]}_{(1)}\partial_{[\mu}\beta_{\nu]}-J^{\mu}_{(1)}\partial_{\mu}\alpha-S^{\mu\alpha\beta}_{(1)}\partial_{\mu}\Omega_{\alpha\beta}+2\Omega_{\alpha\beta}T^{[\alpha\beta]}_{(1)}.
\label{equ15ver1}
\end{align}
Here $\Omega_{\alpha\beta}=\omega_{\alpha\beta}/T=\beta\omega_{\alpha\beta}$.  
Using the condition that for a dissipative system $\partial_{\mu}\mathcal{S}^{\mu}\geq 0$ we can identify $T^{\{\mu\nu\}}_{(1)}$, $T^{[\alpha\beta]}_{(1)}$, $J^{\mu}_{(1)}$, and $S^{\mu\alpha\beta}_{(1)}$ in terms of derivatives of hydrodynamic variables. This is the standard prescription of the Navier-Stokes theory. In this way we can uniquely write $T^{\{\mu\nu\}}_{(1)}$ in terms of $\partial_{\{\mu}\beta_{\nu\}}$, $J^{\mu}_{(1)}$ in terms of $\partial_{\mu}\alpha$, and $S^{\mu\alpha\beta}_{(1)}$ in terms of $\partial_{\mu}\Omega_{\alpha\beta}$.  However a logical contradiction appears for  $T^{[\alpha\beta]}_{(1)}$. From the second term on the right-hand side of the above equation, we can write $T^{[\alpha\beta]}_{(1)}$ in terms of $\partial_{[\mu}\beta_{\nu]}$. But using the last term on the right-hand side of Eq.~\eqref{equ15ver1} we can also write $T^{[\alpha\beta]}_{(1)}$ in terms of $\Omega_{\alpha\beta}$. This is inconsistent with the hydrodynamic gradient expansion scheme, because $T^{[\alpha\beta]}_{(1)}\sim \mathcal{O}(\partial)$, but $\Omega_{\alpha\beta}\sim \mathcal{O}(1)$. One of the ways to remove this logical inconsistency is if we set $T^{[\alpha\beta]}_{(1)}=0$, i.e., up to $\mathcal{O}(\partial)$ there should not be any anti-symmetric part in the energy-momentum tensor. In this case, the divergence of the entropy current boils down to
\begin{align}
\partial_{\mu}\mathcal{S}^{\mu}=T^{\{\mu\nu\}}_{(1)}\partial_{\{\mu}\beta_{\nu\}}-J^{\mu}_{(1)}\partial_{\mu}\alpha-S^{\mu\alpha\beta}_{(1)}\partial_{\mu}\Omega_{\alpha\beta}.
\label{equ16ver1}
\end{align}
To proceed further, we write the a generic decomposition of $T^{\{\alpha\beta\}}_{(1)}$ and  $S^{\mu\alpha\beta}_{(1)}$ in terms of the irreducible tensors which are $\mathcal{O}(\partial)$ in the gradient expansion~\cite{Hattori:2019lfp,Fukushima:2020ucl,Becattini:2011ev} (see Ref.~\cite{Biswas:2023qsw} for a detailed discussion),
\begin{align} 
& T^{\{\alpha\beta\}}_{(1)} = h^{\alpha}u^{\beta}+h^{\beta}u^{\alpha}+\pi^{\alpha\beta}+\Pi\Delta^{\alpha\beta},
\label{equ17ver1}\\
& S^{\mu\alpha\beta}_{(1)}=2u^{[\alpha}\Delta^{\mu\beta]}\Phi+2u^{[\alpha}\tau^{\mu\beta]}_{(s)}+2u^{[\alpha}\tau^{\mu\beta]}_{(a)}+\Theta^{\mu\alpha\beta}.
\label{equ18ver1}
\end{align}
Here, $h^{\mu}$, $\pi^{\mu\nu}$, and $\Pi$ can be identified as heat flux, shear, and bulk viscosity terms, respectively.  Different $\mathcal{O}(\partial)$ currents satisfy the following conditions: $ h^{\mu} u_{\mu}=0, \pi^{\mu\nu}u_{\mu}=0, \pi^{\mu\nu}=\pi^{\nu\mu}, \pi^{\mu}_{~\mu}=0$. 
The $\mathcal{O}(\partial)$ terms appearing in the spin tensor are $\Phi,\tau^{\mu\nu}_{(s)}, \tau^{\mu\nu}_{(a)} $, and $\Theta^{\mu\alpha\beta}$. These currents satisfy the following conditions: $u_{\mu} \tau_{(s)}^{\mu \beta}=0$; $u_{\mu} \tau_{(a)}^{\mu \beta}=0$; $u_{\mu} \Theta^{\mu \alpha \beta}=0$; $u_{\alpha} \Theta^{\mu \alpha \beta}=0$;  $ \tau_{(s)}^{\mu \beta}=\tau_{(s)}^{\beta \mu}$; $\tau_{(a)}^{\mu \beta}=-\tau_{(a)}^{\beta\mu}$; $\Theta^{\mu \alpha \beta}=-\Theta^{\mu \beta \alpha}$. Similar to $\pi^{\mu\nu}$, the term $\tau^{\mu\nu}_{(s)}$ is also traceless, i.e., $\tau^{\mu}_{~~\mu(s)}=0$. Using the explicit expressions for $T^{\{\alpha\beta\}}_{(1)}$, $S^{\mu\alpha\beta}_{(1)}$, in Eq.~\eqref{equ16ver1}, and following the methods discussed in Refs.~\cite{Daher:2022xon,Biswas:2023qsw}  we find, 
\begin{align}
\partial_{\mu}\mathcal{S}^{\mu}= &-\beta h^{\mu}\left(\beta \nabla_{\mu}T-Du_{\mu}\right)+\beta\pi^{\mu\nu}\sigma_{\mu\nu}+\beta\Pi \theta -J^{\mu}_{(1)}\nabla_{\mu}\alpha\nonumber\\
& -2\Phi u^{\alpha}\nabla^{\beta}(\beta\omega_{\alpha\beta})-2\tau^{\mu\beta}_{(s)}u^{\alpha}\Delta^{\gamma\rho}_{\mu\beta}\nabla_{\gamma}(\beta\omega_{\alpha\rho})-2\tau^{\mu\beta}_{(a)}u^{\alpha}\Delta^{[\gamma\rho]}_{[\mu\beta]}\nabla_{\gamma}(\beta\omega_{\alpha\rho})\nonumber\\
& -\Theta_{\mu\alpha\beta}\Delta^{\alpha\delta}\Delta^{\beta\rho}\Delta^{\mu\gamma}\nabla_{\gamma}(\beta\omega_{\delta\rho}),\\
=~& h^{\mu}\left(\frac{n}{\varepsilon+P}\nabla_{\mu}\alpha+\frac{S^{\alpha\beta}}{\varepsilon+P}\nabla_{\mu}\Omega_{\alpha\beta}\right)+\beta\pi^{\mu\nu}\sigma_{\mu\nu}+\beta\Pi \theta -J^{\mu}_{(1)}\nabla_{\mu}\alpha\nonumber\\
& -2\Phi u^{\alpha}\nabla^{\beta}(\beta\omega_{\alpha\beta})-2\tau^{\mu\beta}_{(s)}u^{\alpha}\Delta^{\gamma\rho}_{\mu\beta}\nabla_{\gamma}(\beta\omega_{\alpha\rho})-2\tau^{\mu\beta}_{(a)}u^{\alpha}\Delta^{[\gamma\rho]}_{[\mu\beta]}\nabla_{\gamma}(\beta\omega_{\alpha\rho})\nonumber\\
& -\Theta_{\mu\alpha\beta}\Delta^{\alpha\delta}\Delta^{\beta\rho}\Delta^{\mu\gamma}\nabla_{\gamma}(\beta\omega_{\delta\rho})\nonumber\\
=~& h^{\mu}\frac{S^{\alpha\beta}}{\varepsilon+P}\nabla_{\mu}\Omega_{\alpha\beta}-\mathcal{J}^{\mu}\nabla_{\mu}\alpha+\beta\pi^{\mu\nu}\sigma_{\mu\nu}+\beta\Pi \theta \nonumber\\
& -2\Phi u^{\alpha}\nabla^{\beta}(\beta\omega_{\alpha\beta})-2\tau^{\mu\beta}_{(s)}u^{\alpha}\Delta^{\gamma\rho}_{\mu\beta}\nabla_{\gamma}(\beta\omega_{\alpha\rho})-2\tau^{\mu\beta}_{(a)}u^{\alpha}\Delta^{[\gamma\rho]}_{[\mu\beta]}\nabla_{\gamma}(\beta\omega_{\alpha\rho})\nonumber\\
& -\Theta_{\mu\alpha\beta}\Delta^{\alpha\delta}\Delta^{\beta\rho}\Delta^{\mu\gamma}\nabla_{\gamma}(\beta\omega_{\delta\rho})
\label{equ19ver1}
\end{align}
Here $\mathcal{J}^{\mu} = J^{\mu}_{(1)}-\frac{n}{\varepsilon+P}h^{\mu}$. Imposing the condition that $\partial_{\mu}\mathcal{S}^{\mu}\geq0$ for a dissipative system, we obtain the following relations involving various dissipative currents and first-order derivative of hydrodynamic variables,
\begin{align}
& \Pi=\zeta\theta,\label{equ20ver1}\\
& h^{\mu}=-\kappa_{11}\frac{S^{\alpha\beta}}{\varepsilon+P}\nabla^{\mu}\Omega_{\alpha\beta}-\kappa_{12}\nabla^{\mu}\alpha,\label{equ21ver1}\\
& \pi^{\mu\nu}=2\eta\sigma^{\mu\nu},\label{equ22ver1}\\
& \mathcal{J}^{\mu}=\widetilde{\kappa}_{11}\nabla^{\mu}\alpha+\widetilde{\kappa}_{12}\frac{S^{\alpha\beta}}{\varepsilon+P}\nabla^{\mu}\Omega_{\alpha\beta},\label{equ23ver1}\\
& \Phi=-2\chi_{1} u^{\alpha}\nabla^{\beta}(\beta\omega_{\alpha\beta}),
\label{equ24ver1}\\
& \tau^{\mu\beta}_{(s)}=-2\chi_{2}\Delta^{\mu\beta,\gamma\rho}\nabla_{\gamma}(\beta\omega_{\alpha\rho})u^{\alpha},\label{equ25ver1}\\
& \tau^{\mu\beta}_{(a)}=-2\chi_{3} \Delta^{[\mu\beta][\gamma\rho]}\nabla_{\gamma}(\beta\omega_{\alpha\rho})u^{\alpha},
  \label{equ26ver1}\\
& \Theta^{\mu\alpha\beta}= \chi_4 \Delta^{\delta\alpha}\Delta^{\rho\beta}\Delta^{\gamma\mu}\nabla_{\gamma}(\beta\omega_{\delta\rho})\label{equ27ver1}.
\end{align}
$\eta\geq 0$ and $\zeta\geq 0$ are, respectively, the shear and bulk viscosity coefficients that appear in the standard hydrodynamic frameworks. In the $(h^{\mu},\mathcal{J}^{\mu})$ sector we have four transport coefficients, $\kappa_{11}, \kappa_{12}, \widetilde{\kappa}_{11}$, and $\widetilde{\kappa}_{12}$. $\kappa_{11}$, and $\widetilde{\kappa}_{11}$ are diagonal transport coefficients which relates $h^{\mu}$, and $\mathcal{J}^{\mu}$ with the gradient of the $\beta\omega^{\alpha\beta}$, and $\beta\mu$,  respectively. On the other hand $\kappa_{12}$, $\widetilde{\kappa}_{12}$ are the cross-conductivity transport coefficients. Since we have considered that $\omega^{\alpha\beta}\sim\mathcal{O}(1)$, this automatically implies that $\frac{S^{\alpha\beta}}{\varepsilon+P}\nabla_{\mu}\Omega_{\alpha\beta}\sim \mathcal{O}(\partial)$. Moreover $\nabla_{\mu}\alpha$ is also $\mathcal{O}(\partial)$ term. Hence, both of these terms appear in Eq.~\eqref{equ19ver1}, and transport coefficients related to cross conductivity arise. Such cross-conductivity does not appear if one considers $\omega^{\alpha\beta}\sim\mathcal{O}(\partial)$. In such a case, $\frac{S^{\alpha\beta}}{\varepsilon+P}\nabla_{\mu}\Omega_{\alpha\beta}\sim \mathcal{O}(\partial^2)$ and such terms do not contribute to the entropy production within the framework of first-order spin hydrodynamic framework. This is evident from Eq.~\eqref{equ19ver1} because if we consider  $\omega^{\alpha\beta}\sim\mathcal{O}(\partial)$ then $h^{\mu}\frac{S^{\alpha\beta}}{\varepsilon+P}\nabla_{\mu}\Omega_{\alpha\beta}$ term will be of the order $\mathcal{O}(\partial^3)$ which is one order higher than the term $\mathcal{J}^{\mu}\nabla_{\mu}\alpha$.  By using the Onsager relation, one can show that $\kappa_{12}=\widetilde{\kappa}_{12}$~\cite{Landau_Physical_kinetics}.
Moreover $\partial_{\mu}\mathcal{S}^{\mu}\geq 0$ implies that, $\kappa_{11}\geq 0$, $\widetilde{\kappa}_{11}\geq 0$ and $\kappa_{12}^{2}-\kappa_{11}\widetilde{\kappa}_{11}\leq0$. The transport coefficients $\chi_1\geq 0$, $\chi_2\geq 0$, $\chi_{2}\geq 0$, and $\chi_{4}\geq 0$ are spin-transport coefficients.
Once again, note that if one considers $\omega^{\alpha\beta}\sim\mathcal{O}(\partial)$, then also the spin transport coefficients do not appear within the first order spin hydrodynamic framework~\cite{Hattori:2019ahi,Fukushima:2020ucl,Daher:2022wzf,Daher:2022xon,Biswas:2022bht}. This closes the first-order Navier-Stokes theory of relativistic spin hydrodynamic framework with spin chemical potential leading order in the gradient expansion, i.e., $\omega^{\mu\nu}\sim\mathcal{O}(1)$. 

In the next sections, we explain the method of finding the Kubo formulae for these transport coefficients. 
It is important to note that the first term in Eq.~\eqref{equ19ver1}, i.e., $h^{\mu}S^{\alpha\beta}\nabla_{\mu}\Omega_{\alpha\beta}/(\varepsilon+P)$ introduces some nontriviality in the spin hydrodynamic framework when it comes to the choice of hydrodynamic frames. In the absence of this term, one gets an equivalent hydrodynamic description by setting either $h^{\mu}=0$ or $J^{\mu}_{(1)}=0$. The choice $h^{\mu}=0$ is the Landau frame choice and $J^{\mu}_{(1)}=0$ is the Eckart frame choice. Both of these choices are equivalent in the standard hydrodynamic description. This is because at the level of $\partial_{\mu}\mathcal{S}^{\mu}\geq 0$, we do not remove any term that gives rise to dissipation by choosing either $h^{\mu}=0$, or $J^{\mu}_{(1)}=0$. But in Eq.~\eqref{equ19ver1} $h^{\mu}=0$, and $J^{\mu}_{(1)}=0$ are not equivalent. If we set $h^{\mu}=0$, then we are basically dropping an entropy-producing term originating from the gradient of the spin chemical potential ($\nabla_{\mu}(\beta\omega_{\alpha\beta})$). This situation is certainly different from the situation when we set $J^{\mu}_{(1)}=0$ in the entropy production equation. 
Hence the term  $h^{\mu}S^{\alpha\beta}\nabla_{\mu}\Omega_{\alpha\beta}/(\varepsilon+P)$ indicates to a lack of a Landau-like frame choice. It is also interesting to note that the presence of the spin density and the corresponding chemical potential can give rise to a vector gradient force that preserves parity $S^{\alpha\beta}\nabla_{\mu}\Omega_{\alpha\beta}$.

\subsection{An alternative approach to obtain the spin transport coefficients}
\label{sec2c}
Before we move on to the Kubo relation for various transport coefficients that appear in the first-order spin hydrodynamic frameworks, we first discuss an alternative approach to obtain the spin transport coefficients. In this method, instead of writing $S^{\mu\alpha\beta}_{(1)}$ as given in Eq.~\eqref{equ18ver1}, we prescribe an alternative systematic way to decompose the dissipative part of the spin tensor,
\begin{align}
S_{(1)}^{\mu\alpha\beta} &= \Sigma^{\mu\alpha\beta\eta\gamma\delta}\nabla_{\eta}(\beta\omega_{\gamma\delta})~. \label{eq:dcom_d_SAM}
\end{align}
To construct the projection tensor $\Sigma_{\mu\alpha\beta\eta\gamma\delta}$ one needs to take care of these properties \footnote{Note that $\omega^{\alpha\beta}$ is antisymmetric tensor field that can be decomposed into electric-like and magnetic-like components. If one considers a situation where these vector fields are strong enough to make the system anisotropic, then the decomposition given in Eq.~\eqref{eq:dcom_d_SAM} might have an upper edge than the previous decomposition. This is analogous to the tensor decomposition in the presence of a strong background magnetic field. }, 
\begin{itemize}
\item
The projection tensor containing dissipative currents is orthogonal to $u^{\mu}$, i.e.,
\begin{align}
 u_{\mu}\Sigma^{\mu\alpha\beta\eta\gamma\delta}=0\,.
\end{align}
This is because the dissipative spin current $S^{\mu\alpha\beta}_{(1)}$ is orthogonal to $u^{\mu}$ by construction. 
\item
The projection tensor must be antisymmetric in the second and third indices. It should also be antisymmetric in the last two insides, i.e., $\Sigma^{\mu\alpha\beta\eta\gamma\delta}$ should be antisymmetric in $(\alpha,\beta)$ and $(\gamma,\delta)$,
\begin{align}
    \Sigma^{\mu\alpha\beta\eta\gamma\delta}=\Sigma^{\mu[\alpha\beta]\eta\gamma\delta}=\Sigma^{\mu\alpha\beta\eta[\gamma\delta]}\, ,
\end{align}
\item
The projection tensor should be consistent with the Onsager relation for the time covariance of the microscopic physics,
\begin{align}
    \Sigma^{\mu\alpha\beta\eta\gamma\delta}=\Sigma^{\eta\gamma\delta\mu\alpha\beta}\,.\label{ Onsagercon}
\end{align}
\end{itemize}
The nontrivial task is to find out the irreducible tensor decomposition of $\Sigma^{\mu\alpha\beta\nu\gamma\delta}$ from the available tensors $(u^{\mu},\Delta^{\mu\nu}$ and $\epsilon^{\mu\alpha\beta\gamma})$\footnote{The inclusion of \( \epsilon^{\mu\alpha\beta\gamma} \) is essential for decomposing the projection tensor \( \Sigma^{\mu\alpha\beta\eta\gamma\delta} \) in the case of parity-violating scenarios. However, in the present discussion, we focus on the parity-invariant case. In this context, only one tensor block, \( \mathcal{P}^{\prime\,\mu\alpha\beta\eta\gamma\delta}_{2} \), is spanned by \( \epsilon^{\mu\alpha\beta\gamma} \), as shown in Eq.~\eqref{eq:ortho_basis1}, because it involves two \( \epsilon \)-tensors, which form a parity-invariant block.  }
so that we can identify the dissipative currents uniquely employing the entropy current analysis. In order to achieve the decomposition of $\Sigma^{\mu\alpha\beta\nu\gamma\delta}$, we first define a set of tensor structures preserving the three properties mentioned, which are
\begin{align}
    \mathcal{P}_{1}^{\mu\alpha\beta\eta\gamma\delta}&=\Delta^{\mu[\alpha}\Delta^{\beta][\gamma}
    \Delta^{\delta]\eta},\nonumber\\
    \mathcal{P}_{2}^{\mu\alpha\beta\eta\gamma\delta}&=\Delta^{\mu[\gamma}
    \Delta^{\delta][\alpha}\Delta^{\beta]\eta},\nonumber\\
    \mathcal{P}_{3}^{\mu\alpha\beta\eta\gamma\delta}&=\Delta^{\mu\eta}\Delta^{\gamma[\alpha}
    \Delta^{\beta]\delta},\nonumber\\
    \mathcal{Q}_{1}^{\mu\alpha\beta\eta\gamma\delta}&=\Delta^{\mu[\alpha}u^{\beta]}
    \Delta^{\eta[\gamma}u^{\delta]},\nonumber\\
    \mathcal{Q}_{2}^{\mu\alpha\beta\eta\gamma\delta}&=\Delta^{\mu\eta}\Delta^{[\alpha[\gamma}u^{\delta]}
    u^{\beta]},\nonumber\\
    \mathcal{Q}_{3}^{\mu\alpha\beta\eta\gamma\delta}&=\Delta^{\mu[\gamma}u^{\delta]}\Delta^{\eta[\alpha}
    u^{\beta]}.
    \label{equ29new}
\end{align}
It is important to note that $\mathcal{P}_{i}$'s and $\mathcal{Q}_{j}$'s as defined in Eq.~\eqref{equ29new} do not constitute a set of an orthogonal set of basis. However, such an orthogonal set of bases can be obtained by taking linear combinations of $\mathcal{P}_{i}$'s and $\mathcal{Q}_{j}$'s. Moreover, the second basis of Eq.~\eqref{equ29new} can be replaced in terms of Levi-Civita tensors because of the identity, 
\begin{align}
\varepsilon_{\mu\alpha\beta}\varepsilon_{\eta\gamma\delta}= &  u^{\lambda} u^{\rho} \epsilon_{\lambda\mu \alpha \beta} \epsilon_{\rho \eta\gamma\delta}=u^{\lambda} u^{\rho} \left|\begin{array}{cccc}
g_{\lambda \rho} & g_{\lambda \eta} & g_{\lambda \gamma} & g_{\lambda \delta} \\
g_{\mu \rho} & g_{\mu \eta} & g_{\mu \gamma} & g_{\mu \delta}  \\
g_{\alpha \rho} & g_{\alpha \eta} & g_{\alpha \gamma} & g_{\alpha \delta}  \\
g_{\beta \rho} & g_{\beta \eta} & g_{\beta \gamma} & g_{\beta \delta} 
\end{array}\right| \label{epep}\\
&=4\Bigg(\mathcal{P}_{2,\mu\alpha\beta\eta\gamma\delta}+\frac{1}{2}\mathcal{P}_{3,\mu\alpha\beta\eta\gamma\delta} \Bigg).\nonumber
\end{align}
We obtained the following orthogonal set of basis in terms of $\mathcal{P}_{i}$'s and $\mathcal{Q}_{j}$'s:
\begin{align}
    \mathcal{P}_{1}^{\prime\mu\alpha\beta\eta\gamma\delta} &=\mathcal{P}_{1}^{\mu\alpha\beta\eta\gamma\delta}~,
    \nonumber\\
    \mathcal{P}_{2}^{\prime\mu\alpha\beta\eta\gamma\delta} &=\frac{4}{6}\Bigg(\mathcal{P}_{2,\mu\alpha\beta\eta\gamma\delta}+\frac{1}{2}\mathcal{P}_{3,\mu\alpha\beta\eta\gamma\delta} \Bigg)=\frac{u_{\lambda}}{6}\epsilon^{\lambda\mu\alpha\beta}u_{\nu}\epsilon^{\nu\eta\gamma\delta}=\frac{1}{6}\varepsilon^{\mu\alpha\beta}\varepsilon^{\eta\gamma\delta}~,
    \nonumber\\
    \mathcal{P}_{3}^{\prime\mu\alpha\beta\eta\gamma\delta} &=\mathcal{P}_{3}^{\mu\alpha\beta\eta\gamma\delta}+\mathcal{P}_{1}^{\mu\alpha\beta\eta\gamma\delta}+\mathcal{P}_{2}^{\prime\,\mu\alpha\beta\eta\gamma\delta},
    \nonumber\\
    \mathcal{Q}_{1}^{\prime\mu\alpha\beta\eta\gamma\delta} &=\mathcal{Q}_{1}^{\mu\alpha\beta\eta\gamma\delta}~,
    \nonumber\\
    \mathcal{Q}_{2}^{\prime\mu\alpha\beta\eta\gamma\delta} &=\mathcal{Q}_{2}^{\mu\alpha\beta\eta\gamma\delta}+\mathcal{Q}_{3}^{\mu\alpha\beta\eta\gamma\delta}-\frac{2}{3}\mathcal{Q}_{1}^{\mu\alpha\beta\eta\gamma\delta}~,
    \nonumber\\
    \mathcal{Q}_{3}^{\prime\mu\alpha\beta\eta\gamma\delta} &=\mathcal{Q}_{2}^{\mu\alpha\beta\eta\gamma\delta}-\mathcal{Q}_{3}^{\mu\alpha\beta\eta\gamma\delta}.
 \label{eq:ortho_basis1}
\end{align}
Now, one can explicitly check that $\mathcal{P}_{i}$'s and $\mathcal{Q}_{j}$'s satisfy orthogonality conditions:
\begin{align}
 \mathcal{P}^{\prime\mu\alpha\beta\eta\gamma\delta}_{i}\mathcal{P}_{j\,\mu\alpha\beta\eta\gamma\delta}^{\prime} &=0~,
    \\
    \mathcal{Q}^{\prime\mu\alpha\beta\eta\gamma\delta}_{i}\mathcal{P}_{j\,\mu\alpha\beta\eta\gamma\delta}^{\prime} &=0~,
    \\
    \mathcal{Q}^{\prime\mu\alpha\beta\eta\gamma\delta}_{i}\mathcal{Q}_{j\,\mu\alpha\beta\eta\gamma\delta}^{\prime} &=0~,
    \label{eq:ortho_condition}
\end{align}
for all $i, j$ and for $i\neq j$. Once we have identified the orthogonal set of tensors, we can decompose $\Sigma^{\mu\alpha\beta\eta\gamma\delta}$ in terms of orthogonal bases $(\mathcal{P}^{\prime\,\mu\alpha\beta\eta\gamma\delta}_{i},\mathcal{Q}^{\prime\,\mu\alpha\beta\eta\gamma\delta}_{i})$ with the set of coefficients $(\Sigma_{i},\Lambda_{i})$,
\begin{align}
    \Sigma^{\mu\alpha\beta\eta\gamma\delta}
    &=\Sigma_{1}\mathcal{P}_{1}^{\prime\mu\alpha\beta\eta\gamma\delta}+\Sigma_{2}\mathcal{P}_{2}^{\prime\mu\alpha\beta\eta\gamma\delta}+\Sigma_{3}\mathcal{P}_{3}^{\prime\mu\alpha\beta\eta\gamma\delta}+\Lambda_{1}\mathcal{Q}_{1}^{\prime\mu\alpha\beta\eta\gamma\delta}+\Lambda_{2}\mathcal{Q}_{2}^{\prime\mu\alpha\beta\eta\gamma\delta}+\Lambda_{3}\mathcal{Q}_{3}^{\prime\mu\alpha\beta\eta\gamma\delta}\\
    &=\big(\Sigma_{1}+\Sigma_{3}\big)\mathcal{P}_{1}^{\prime\mu\alpha\beta\eta\gamma\delta}+\big(\Sigma_{2}+\Sigma_{3}\big)\mathcal{P}_{2}^{\prime\mu\alpha\beta\eta\gamma\delta}+\Sigma_{3}\mathcal{P}_{3}^{\mu\alpha\beta\eta\gamma\delta}+\Lambda_{1}\mathcal{Q}_{1}^{\prime\mu\alpha\beta\eta\gamma\delta}\nonumber\\
    &+\Lambda_{2}\mathcal{Q}_{2}^{\prime\mu\alpha\beta\eta\gamma\delta}+\Lambda_{3}\mathcal{Q}_{3}^{\prime\mu\alpha\beta\eta\gamma\delta}.\label{equ34new}
\end{align}
If we consider the above expression of $\Sigma^{\mu\alpha\beta\eta\gamma\delta}$ then we can identify the contribution of $S_{(1)}^{\mu\alpha\beta}$ that appears in the divergence of the entropy four current, i.e., 
\begin{align}
\partial_{\mu}\mathcal{S}^{\mu}|_{S_{(1)}^{\mu\alpha\beta}} = & -(\Sigma_1+\Sigma_3) \nabla_{\mu}(\beta \omega_{\alpha\beta})\mathcal{P}_{1}^{\prime\mu\alpha\beta\eta\gamma\delta} \nabla_{\eta}(\beta \omega_{\gamma\delta})-\left(\Sigma_2+\Sigma_3\right) \nabla_{\mu}(\beta \omega_{\alpha\beta})\mathcal{P}_{2}^{\prime\mu\alpha\beta\eta\gamma\delta} \nabla_{\eta}(\beta \omega_{\gamma\delta})\nonumber\\
& -\Sigma_3 \nabla_{\mu}(\beta \omega_{\alpha\beta})\mathcal{P}_{3}^{\prime\mu\alpha\beta\eta\gamma\delta} \nabla_{\eta}(\beta \omega_{\gamma\delta})
 -\Lambda_1 \nabla_{\mu}(\beta \omega_{\alpha\beta})\mathcal{Q}_{1}^{\prime\mu\alpha\beta\eta\gamma\delta} \nabla_{\eta}(\beta \omega_{\gamma\delta})\nonumber\\
 &  -\Lambda_2 \nabla_{\mu}(\beta \omega_{\alpha\beta})\mathcal{Q}_{2}^{\prime\mu\alpha\beta\eta\gamma\delta} \nabla_{\eta}(\beta \omega_{\gamma\delta})
  -\Lambda_3 \nabla_{\mu}(\beta \omega_{\alpha\beta})\mathcal{Q}_{3}^{\prime\mu\alpha\beta\eta\gamma\delta} \nabla_{\eta}(\beta \omega_{\gamma\delta}).
\end{align}  

After performing more simplification, one can show that, 
\begin{align}
&(i) -(\Sigma_1+\Sigma_3) \nabla_{\mu}(\beta \omega_{\alpha\beta})\mathcal{P}_{1}^{\prime\mu\alpha\beta\eta\gamma\delta} \nabla_{\eta}(\beta \omega_{\gamma\delta})=(\Sigma_1+\Sigma_3) \Phi_{\beta\perp} \Phi^{\beta\perp},\label{equ36new}\\
&(ii) -\left(\Sigma_2+\Sigma_3\right) \nabla_{\mu}(\beta \omega_{\alpha\beta})\mathcal{P}_{2}^{\prime\mu\alpha\beta\eta\gamma\delta} \nabla_{\eta}(\beta \omega_{\gamma\delta}) =-6\left(\Sigma_2+\Sigma_3\right) \varphi^2, \label{equ37new}\\
&(iii) -\Sigma_3 \nabla_{\mu}(\beta \omega_{\alpha\beta})\mathcal{P}_{3}^{\mu\alpha\beta\eta\gamma\delta} \nabla_{\eta}(\beta \omega_{\gamma\delta})= -\Sigma_3 \Phi^{\eta\gamma\delta} \Phi_{\eta\gamma\delta},\label{equ38new}\\
&(iv)-\Lambda_1 \nabla_{\mu}(\beta \omega_{\alpha\beta})\mathcal{Q}_{1}^{\prime\mu\alpha\beta\eta\gamma\delta} \nabla_{\eta}(\beta \omega_{\gamma\delta})= -\Lambda_1 \Phi_{||}^2, \label{equ39new}\\
&(v) -\Lambda_2 \nabla_{\mu}(\beta \omega_{\alpha\beta})\mathcal{Q}_{2}^{\prime\mu\alpha\beta\eta\gamma\delta} \nabla_{\eta}(\beta \omega_{\gamma\delta}) =-\frac{\Lambda_2}{2} \Gamma^{\mu\alpha}_{(s)}\Gamma_{(s)\mu\alpha}, \label{equ40new}\\
&(vi)-\Lambda_3 \nabla_{\mu}(\beta \omega_{\alpha\beta})\mathcal{Q}_{3}^{\prime\mu\alpha\beta\eta\gamma\delta} \nabla_{\eta}(\beta \omega_{\gamma\delta}) = -\frac{\Lambda_3}{2} \Gamma^{\mu\alpha}_{(a)}\Gamma_{(a)\mu\alpha}. \label{equ41new}
\end{align}
Here, $\Phi^{\mu\alpha\beta}=\Delta^{\alpha[\gamma}\Delta^{\delta]\beta}\nabla^{\mu}(\beta\omega_{\gamma\delta})$,\,$\Phi^{\beta}=\nabla_{\mu}(\beta\omega^{\mu\beta})=\Phi_{\perp}^{\beta}+\Phi_{||}^{\beta}$, $\varphi=\frac{1}{6}\varepsilon^{\eta\gamma\delta}\nabla_{\eta}(\beta\omega_{\gamma\delta})$ and $\Gamma^{\mu\alpha}=\nabla^{\mu}\big(\beta\omega^{\alpha\beta}\big)u_{\beta}$. Moreover, $\Gamma_{(s)}^{\mu\alpha}$ and $\Gamma_{(a)}^{\mu\alpha}$ are the traceless symmetric and antisymmetric part of $\Gamma^{\mu\alpha}$, on the other hand, $\Phi^{\beta}_{||}$ contains the symmetric trace part of $\Gamma^{\mu\alpha}$. A noteworthy point is that every term mentioned in Eqs.~\eqref{equ36new}-\eqref{equ41new} occurs in the divergence of entropy due to the spin contribution. Each term contains a square-completed term.
The sign of different coefficients ($\Sigma_i,\Lambda_i$) that appear in Eq.~\eqref{equ34new} can be obtained by demanding the positivity of the entropy divergence in the spin part of the total entropy current, i.e., $\partial_{\mu}\mathcal{S}^{\mu}|_{S_{(1)}^{\mu\alpha\beta}}\geq 0$ gives
\begin{align}
    &\Sigma_{1}+\Sigma_{3}\leq0,\,\Sigma_{2}+\Sigma_{3}\leq0,\,\Sigma_{3}\geq0,\,\nonumber\\
    &\Lambda_{1}\leq0,\,\Lambda_{2}\leq0,\,\Lambda_{3}\leq0.\,
\end{align}
Note that one can extract the transport coefficients from the six rank tensor $ \Sigma^{\mu\alpha\beta\eta\gamma\delta}$ by contracting it with the orthogonal projection basis $\mathcal{P}^{\prime}_{i}$'s and $\mathcal{Q}^{\prime}_{j}$'s,
\begin{align}
    & \Sigma^{\mu\alpha\beta\eta\gamma\delta}\mathcal{P}^{\prime}_{1\mu\alpha\beta\eta\gamma\delta}=3\Sigma_{1},\nonumber\\
    & \Sigma^{\mu\alpha\beta\eta\gamma\delta}\mathcal{P}^{\prime}_{2\mu\alpha\beta\eta\gamma\delta}=\Sigma_{2}, \nonumber\\
    & \Sigma^{\mu\alpha\beta\eta\gamma\delta}\mathcal{P}^{\prime}_{3\mu\alpha\beta\eta\gamma\delta}=5\Sigma_{3},\nonumber\\
    & \Sigma^{\mu\alpha\beta\eta\gamma\delta}\mathcal{Q}^{\prime}_{1\mu\alpha\beta\eta\gamma\delta}=\frac{9}{4}\Lambda_{1}, \nonumber   \\
    & \Sigma^{\mu\alpha\beta\eta\gamma\delta}\mathcal{Q}^{\prime}_{2\mu\alpha\beta\eta\gamma\delta}=5\Lambda_{2},\nonumber\\
 & \Sigma^{\mu\alpha\beta\eta\gamma\delta}\mathcal{Q}^{\prime}_{3\mu\alpha\beta\eta\gamma\delta}=3\Lambda_{3}.
    \label{equ49new}
\end{align}
Eventually, one can write the dissipative spin current as a sum of six independent dissipative components,
\begin{align}
     S_{(1)}^{\mu\alpha\beta}=-\big(\Sigma_{1}+\Sigma_{3}\big)\Delta^{\mu[\alpha}\Phi^{\beta]}_{\perp}+\big(\Sigma_{2}+\Sigma_{3}\big)\varepsilon^{\mu\alpha\beta}\varphi+\Sigma_{3}\Phi^{\mu\alpha\beta}+\Lambda_{1}\Delta^{\mu[\alpha}\Phi^{\beta]}_{||}+\Lambda_{2}\Gamma_{s}^{\mu[\alpha}u^{\beta]}+\Lambda_{3}\Gamma_{a}^{\mu[\alpha}u^{\beta]}\,.
     \label{equ54new}
\end{align}
Some comparison between the decomposition of the dissipative spin current $S^{\mu\alpha\beta}_{(1)}$ as given in Eqs.~\eqref{equ18ver1} and \eqref{equ54new} is in order here. The irreducible dissipative currents that appear in Eqs.~\eqref{equ18ver1} are $\Phi$, $\tau^{\mu\beta}_{(s)}$, $\tau^{\mu\beta}_{(a)}$, and $\Theta_{\mu\alpha\beta}$. It is important to note that $\Phi$, $\tau^{\mu\beta}_{(s)}$, $\tau^{\mu\beta}_{(a)}$, and $\Theta_{\mu\alpha\beta}$ have respectively one, five, three, and nine independent components or degrees of freedom which adds up to eighteen independent components which represents the total number of the degrees of freedom of $S^{\mu\alpha\beta}_{(1)}$. On the other hand, in Eq.~\eqref{equ54new} the irreducible dissipative currents are $\Phi_{||}=\Phi^{\beta}_{||}u_{\beta}$, $\Gamma^{\mu\alpha}_{(s)}$, $\Gamma^{\mu\alpha}_{(a)}$, $\Phi^{\beta}_{\perp}$, $\varphi$, and $\Phi^{\mu\alpha\beta}$. Together, these dissipative currents also add up to eighteen independent components of $S^{\mu\alpha\beta}_{(1)}$.
The equivalency between the two representations of $S^{\mu\alpha\beta}_{(1)}$ can be summarized as $\Phi \equiv \Phi_{||}$, $\tau^{\mu\beta}_{(s)}\equiv \Gamma^{\mu\beta}_{(s)}$, $\tau^{\mu\beta}_{(a)}\equiv \Gamma^{\mu\beta}_{(a)}$, and $\Phi^{\mu\alpha\beta}\equiv\Theta^{\mu\alpha\beta}$~\footnote{Note that these are only equivalence relation between two decompositions. These tensors are not equal because they have different dimensions, e.g., $\Phi$ and  $\Phi_{||}$ have different mass dimensions. This is because, in Eqs.~\eqref{equ18ver1}, transport coefficients do not appear explicitly along with the dissipative currents. But in Eq.~\eqref{equ54new} transport coefficients appear explicitly.}. However, comparing Eq.~\eqref{equ18ver1} with Eq.\eqref{equ54new}, one can identify that two extra terms (first and second terms) are present in Eq.~\eqref{equ54new}, which represents the trace and antisymmetric part of $\Phi^{\mu\alpha\beta}$ (antisymmetric with respect to the
first two indices $(\mu,\alpha)$). The first and the second terms appearing in Eq.~\eqref{equ54new} contain three and one\footnote{ Anti-symmetry under the exchange of $(\mu\leftrightarrow\alpha)$ together with anti-symmetric pair of last two indices $(\alpha,\beta)$ makes it a totally anti-symmetric object, which is represented by $\varepsilon^{\mu\alpha\beta}$ and the number of the independent component becomes one.   } independent degrees of freedom, respectively. Other five\footnote{The projection of $\nabla^{\eta}(\beta\omega^{\gamma\delta})$ along $\mathcal{P}^{\prime\,\mu\alpha\beta\eta\gamma\delta}_{2}$ is traceless with respect to $(\mu,\alpha)$ but anti-symmetric with respect to $(\alpha,\beta)$. This can be verified from  $\mathcal{P}^{\prime\,~~\mu\beta\eta\gamma\delta}_{2,\mu}=0$. } are spanned by $\mathcal{P}^{\prime}_{3\,\mu\alpha\beta\eta\gamma\delta}$ or equivalently hides in the third term with the presence of $\Phi^{\mu\alpha\beta}$. In other words, the latter representation, as in  Eq.~\eqref{equ54new}, first and second terms lifts off the degeneracy of $\Theta^{\mu\alpha\beta}$. Therefore $\Delta^{\mu[\alpha} \Phi^{\beta]}_{\perp}$, $\varepsilon^{\mu\alpha\beta}\varphi$, and the rest of the degrees of freedom in $\Phi^{\mu\alpha\beta}$ account for a total of nine independent components of $\Theta^{\mu\alpha\beta}$.

So far, we have discussed the possible tensor decompositions of various dissipative currents appearing in the energy-momentum tensor and the spin tensor. Moreover, we have also identified the possible transport coefficient that gives rise to dissipation in the system due to various thermodynamic forces. Now, we move on to the microscopic description to obtain the Kubo relation of these transport coefficients in terms of the corresponding current correlators. Furthermore, we have two tensor decompositions of the dissipative part of the spin tensor. Different transport coefficients are also associated with these two tensor decompositions. In the subsequent sections we also show the equivalence of these two decompositions by showing the relations among the independent spin transport coefficients appearing in these two descriptions.

\section{Non-equilibrium statistical operator and Kubo relations}
\label{sec3}
The non-equilibrium statistical operators (NESO) approach pioneered by Zubarev allows
one to develop a statistical theory of irreversible processes. In this approach, 
one considers a statistical ensemble representing the macroscopic
state of the system in a non-equilibrium state. We consider a system in the hydrodynamic regime
characterized by local thermodynamic parameters, e.g., temperature ($T$), chemical potential ($\mu$), and spin chemical potential ($\omega^{\alpha\beta}$).
In a relativistic system, the non-equilibrium statistical operator (NESO) is defined as~\cite{alma991021547569703276,Huang:2011dc,Hu:2021lnx,Hosoya:1983id}
\begin{align}
\widehat{\rho}(t)=\frac{1}{Q}\exp\left[-\int d^3x~\widehat{Z}\left(\vec{x},t\right)\right],
\label{equ28ver1}
\end{align}
The factor $Q$ is the normalization so that $\Tr\widehat{\rho}(t)=1$,
\begin{align}
Q=\Tr \exp\left[-\int d^3x~\widehat{Z}\left(\vec{x},t\right)\right].
\label{equ29ver1}
\end{align}
The operator $\widehat{Z}(\vec{x},t)$ is defined as,
\begin{align}
\widehat{Z}(\vec{x},t)=\epsilon \int _{-\infty}^t~dt^{\prime}~e^{\epsilon(t^{\prime}-t)}\bigg[\beta^{\nu}(\vec{x},t^{\prime})\widehat{T}_{0\nu}(\vec{x},t^{\prime})-\alpha (\vec{x},t^{\prime})\widehat{J}^0(\vec{x},t^{\prime})-\Omega_{\rho\sigma}(\vec{x},t^{\prime})\widehat{S}^{0\rho\sigma}(\vec{x},t^{\prime})\bigg],
\label{equ30ver1}
\end{align}
with $\epsilon\rightarrow +0$. Here, the Lagrange multipliers appearing in Eq.~\eqref{equ30ver1} are 
\begin{align}
& \beta^{\nu}(\vec{x},t^{})=\beta(\vec{x},t^{})u^{\nu}(\vec{x},t^{}),\label{equ31ver1}\\
& \alpha(\vec{x},t^{})=\beta(\vec{x},t^{})\mu(\vec{x},t^{}),\label{equ32ver1}\\
& \Omega^{\mu\nu}(\vec{x},t^{})=\beta(\vec{x},t^{})\omega^{\mu\nu}(\vec{x},t^{}).\label{equ33ver1}
\end{align}
Physically, $\beta(\vec{x},t^{})$, $\mu(\vec{x},t^{})$, $u^{\nu}(\vec{x},t^{})$, and $\omega^{\mu\nu}(\vec{x},t^{})$ represents the inverse of local equilibrium temperature ($T(\vec{x},t^{})$), chemical potential, fluid flow four-velocity, and spin chemical potential, respectively. The
operators $\widehat{T}^{\mu\nu}$, $\widehat{J}^{\mu}$, and $\widehat{S}^{\mu\alpha\beta}$ are energy-momentum tensor, conserved four current, and spin tensor satisfying the local conservation laws,
\begin{align}
\partial_{\mu}\widehat{T}^{\mu\nu}(\vec{x},t^{})=0; ~~\partial_{\mu}\widehat{J}^{\mu}(\vec{x},t^{})=0; ~~\partial_{\lambda}\widehat{S}^{\lambda\mu\nu}(\vec{x},t^{})=0.\label{equ34ver1}
\end{align}
Using the spin hydrodynamic equations, it can be shown that (see Appendix~\ref{appenC} for details), 
\begin{align}
\int d^3x ~\widehat{Z}(\vec{x},t)=\widehat{A}(t)-\widehat{B}(t),
\label{equ35ver1}
\end{align}
here $\widehat{A}(t)$, and $\widehat{B}(t)$ are, 
\begin{align}
\widehat{A}(t) =\int d^3x\bigg[\beta^{\nu}(\vec{x},t^{})\widehat{T}_{0\nu}(\vec{x},t^{})-\alpha (\vec{x},t^{})\widehat{J}^0(\vec{x},t^{})-\Omega_{\rho\sigma}(\vec{x},t^{})\widehat{S}^{0\rho\sigma}(\vec{x},t^{})\bigg],
\label{equ36ver1}
\end{align}
\begin{align}
\widehat{B}(t) & = \int d^3x~\int _{-\infty}^t~dt^{\prime}~e^{\epsilon(t^{\prime}-t)}\bigg[\partial_{\mu}\beta_{\nu}(\vec{x},t^{\prime})\widehat{T}^{\mu\nu}_{}(\vec{x},t^{\prime})-\partial_{\mu}\Omega_{\rho\sigma}(\vec{x},t^{\prime})\widehat{S}^{\mu\rho\sigma}_{}(\vec{x},t^{\prime})-\partial_{\mu}\alpha(\vec{x},t^{\prime})\widehat{J}^{\mu}(\vec{x},t^{\prime})\bigg]\nonumber\\
& = \int d^3x~\int _{-\infty}^t~dt^{\prime}~e^{\epsilon(t^{\prime}-t)}~\widehat{C}(\vec{x},t^{\prime}).
\label{equ37ver1}
\end{align}
In terms of $\widehat{A}(t)$, and $\widehat{B}(t)$ the non-equilibrium statistical operators (NESO) becomes~\cite{alma991021547569703276,Huang:2011dc}, 
\begin{align}
\widehat{\rho}(t)=\frac{1}{Q}\exp\left(-\widehat{A}+\widehat{B}\right), \text{with},~~~Q=\Tr \exp\left(-\widehat{A}+\widehat{B}\right). 
\label{equ38ver1}
\end{align}
In the operator $\widehat{B}(t)$ the gradient of local thermodynamic quantities, i.e., $\partial_{\mu}\beta_{\nu}(\vec{x},t^{})$, $\partial_{\mu}\alpha(\vec{x},t^{})$, and $\partial_{\mu}\Omega_{\rho\sigma}(\vec{x},t^{})$ appear. These gradients of local thermodynamic quantities can be considered as \textit{thermodynamic forces}. In equilibrium, we expect these thermodynamic forces to vanish. Therefore, it is natural to identify the operator $\widehat{B}$ as the non-equilibrium part of the statistical operator, which vanishes in equilibrium. Naturally, one can identify $\widehat{A}$ as the equilibrium part, and we may define the local equilibrium statistical operator as~\cite{alma991021547569703276,Huang:2011dc} 
\begin{align}
\widehat{\rho}_{l}(t)=\frac{1}{Q_l}\exp\left(-\widehat{A}\right), \text{with},~~~Q_l=\Tr \exp\left(-\widehat{A}\right).
\end{align}
In principle, it is not straightforward to obtain a simplified closed-form expression of the statistical operator for generic $\widehat{A}$ and $\widehat{B}$. However, the situation becomes simplified for the linear response theory, where the relations between the thermodynamic forces and irreversible currents are linear. In this case, we consider small perturbations from the local equilibrium state, and the thermodynamic gradients or the thermodynamic forces are sufficiently small. In this regime, we can write the non-equilibrium statistical ($\widehat{\rho}(t)$) operator by considering only the linear terms in perturbation (see Ref.~\cite{alma991021547569703276} and Appendix~\ref{appenD} for details),
\begin{align}
\widehat{\rho}(t)& \simeq \bigg[1+\int_0^1 d\tau \bigg\{e^{-\tau\widehat{A}}\widehat{B}e^{\tau\widehat{A}}-\langle e^{-\tau\widehat{A}}\widehat{B}e^{\tau\widehat{A}}\rangle_l\bigg\}\bigg]\widehat{\rho}_l\nonumber\\
& \simeq \bigg[1+\int_0^1 d\tau \bigg\{e^{-\tau\widehat{A}}\widehat{B}e^{\tau\widehat{A}}-\langle \widehat{B}_{\tau}\rangle_l\bigg\}\bigg]\widehat{\rho}_l,
\end{align}
here $\langle \mathcal{O}\rangle_l=\Tr\left(\widehat{\rho}_l\mathcal{O}\right)$ 
is the statistical average of the operator $\mathcal{O}$ for the local equilibrium operator. Once we have identified the out-of-equilibrium density operator in terms of the local equilibrium operator, we can now calculate the statistical average of the operators corresponding to the microscopic current, i.e.,  $\widehat{T}^{\mu\nu}(\vec{x},t)$, $ \widehat{J}^{\mu}(\vec{x},t)$, and  $\widehat{S}^{\mu\alpha\beta}(\vec{x},t)$. 
The energy-momentum tensor averaged over the non-equilibrium distribution,
\begin{align}
\langle \widehat{T}^{\mu\nu}(\vec{x},t)\rangle & = \Tr\left(\widehat{\rho}(t)\widehat{T}^{\mu\nu}(\vec{x},t)\right)\nonumber\\
& =  \Tr\left(\bigg[1+\int_0^1 d\tau \bigg\{e^{-\tau\widehat{A}}\widehat{B}e^{\tau\widehat{A}}-\langle \widehat{B}_{\tau}\rangle_l\bigg\}\bigg]\widehat{\rho}_l~\widehat{T}^{\mu\nu}(\vec{x},t)\right)\nonumber\\
& =  \Tr\left(\widehat{\rho}_l(t)\widehat{T}^{\mu\nu}(\vec{x},t)\right) + \Tr\left(\int_0^1 d\tau \bigg\{e^{-\tau\widehat{A}}\widehat{B}e^{\tau\widehat{A}}-\langle \widehat{B}_{\tau}\rangle_l\bigg\}\widehat{\rho}_l~\widehat{T}^{\mu\nu}(\vec{x},t)\right)\nonumber\\
& = \langle \widehat{T}^{\mu\nu}(\vec{x},t)\rangle_l+\delta \langle \widehat{T}^{\mu\nu}(\vec{x},t)\rangle,
\end{align}
here, 
\begin{align}
\delta \langle \widehat{T}^{\mu\nu} (\vec{x},t)\rangle = \Tr\left(\int_0^1 d\tau \bigg\{e^{-\tau\widehat{A}}\widehat{B}e^{\tau\widehat{A}}-\langle \widehat{B}_{\tau}\rangle_l\bigg\}\widehat{\rho}_l~\widehat{T}^{\mu\nu}(\vec{x},t)\right).
\end{align}
$\langle \widehat{T}^{\mu\nu}\rangle_l$
represents the statistical average of the energy-momentum tensor for the local equilibrium density operator, and $\delta \langle \widehat{T}^{\mu\nu}\rangle$ is the deviation of the averaged energy-momentum from the local equilibrium. Similarly, we can write the statistical average of the number-current operator and the spin-tensor operator,
\begin{align}
& \langle \widehat{J}^{\mu}(\vec{x},t)\rangle  = \langle \widehat{J}^{\mu}(\vec{x},t)\rangle_l+\delta \langle \widehat{J}^{\mu}(\vec{x},t)\rangle, \\
& \langle \widehat{S}^{\mu\alpha\beta}(\vec{x},t)\rangle = \langle \widehat{S}^{\mu\alpha\beta}(\vec{x},t)\rangle_l+\delta \langle \widehat{S}^{\mu\alpha\beta}(\vec{x},t)\rangle,
\end{align} 
here,
\begin{align}
& \langle \widehat{J}^{\mu}(\vec{x},t)\rangle_l = \Tr\left(\widehat{\rho}_l(t)\widehat{J}^{\mu}(\vec{x},t)\right),\\
&  \langle \widehat{S}^{\mu\alpha\beta}(\vec{x},t)\rangle_l = \Tr\left(\widehat{\rho}_l(t)\widehat{S}^{\mu\alpha\beta}(\vec{x},t)\right),
\end{align}
and,
\begin{align}
& \delta\langle \widehat{J}^{\mu}(\vec{x},t)\rangle = \Tr\left(\int_0^1 d\tau \bigg\{e^{-\tau\widehat{A}}\widehat{B}e^{\tau\widehat{A}}-\langle \widehat{B}_{\tau}\rangle_l\bigg\}\widehat{\rho}_l~\widehat{J}^{\mu}(\vec{x},t)\right),\\
& \delta\langle \widehat{S}^{\mu\alpha\beta}(\vec{x},t)\rangle = \Tr\left(\int_0^1 d\tau \bigg\{e^{-\tau\widehat{A}}\widehat{B}e^{\tau\widehat{A}}-\langle \widehat{B}_{\tau}\rangle_l\bigg\}\widehat{\rho}_l~\widehat{S}^{\mu\alpha\beta}(\vec{x},t)\right).
\end{align}
Using the expression of $\widehat{B}(t)$ as given in Eq.~\eqref{equ37ver1} we can write, 
\begin{align}
\delta\langle \widehat{T}^{\mu\nu}(\vec{x},t)\rangle & = \int d^3x^{\prime}\int_{-\infty}^t dt^{\prime}~e^{\epsilon(t^{\prime}-t)} \int_0^1 d\tau \bigg\langle \widehat{T}^{\mu\nu}(\vec{x},t)\bigg\{e^{-\tau\widehat{A}}\widehat{C}(\vec{x}^{\prime},t^{\prime})e^{\tau\widehat{A}}-\langle\widehat{C}(\vec{x}^{\prime},t^{\prime})_{\tau}\rangle_l\bigg\}\bigg\rangle_l\nonumber\\
& = \int d^3x^{\prime}\int_{-\infty}^t dt^{\prime}~e^{\epsilon(t^{\prime}-t)}\bigg(\widehat{T}^{\mu\nu}(\vec{x},t),\widehat{C}^{}(\vec{x}^{\prime},t^{\prime})\bigg)_l\label{delT}
\end{align}
In the above equation, we have used the following shorthand notation,
\begin{align}
\bigg(\widehat{X}(\vec{x},t),\widehat{Y}^{}(\vec{x}^{\prime},t^{\prime})\bigg)_l = \int_0^1 d\tau \bigg\langle \widehat{X}^{}(\vec{x},t)\bigg\{e^{-\tau\widehat{A}}\widehat{Y}(\vec{x}^{\prime},t^{\prime})e^{\tau\widehat{A}}-\langle\widehat{Y}(\vec{x}^{\prime},t^{\prime})_{\tau}\rangle_l\bigg\}\bigg\rangle_l.
\end{align}
Using the same methodology, it can be shown that 
\begin{align}
\delta\langle \widehat{J}^{\mu}(\vec{x},t)\rangle & = \int d^3x^{\prime}\int_{-\infty}^t dt^{\prime}~e^{\epsilon(t^{\prime}-t)}\bigg(\widehat{J}^{\mu}(\vec{x},t),\widehat{C}^{}(\vec{x}^{\prime},t^{\prime})\bigg)_l,\label{delJ}
\end{align}
and 
\begin{align}
\delta\langle \widehat{S}^{\mu\alpha\beta}(\vec{x},t)\rangle & = \int d^3x^{\prime}\int_{-\infty}^t dt^{\prime}~e^{\epsilon(t^{\prime}-t)}\bigg(\widehat{S}^{\mu\alpha\beta}(\vec{x},t),\widehat{C}^{}(\vec{x}^{\prime},t^{\prime})\bigg)_l.\label{delS}
\end{align}
The statistical average of the microscopic currents, i.e., $\widehat{T}^{\alpha\beta}(\vec{x},t) $, $\widehat{J}^{\mu}(\vec{x},t)$, and $\widehat{S}^{\mu\alpha\beta}(\vec{x},t)$ obtained using the statistical operator $\widehat{\rho}_l$, i.e., $\langle \widehat{T}^{\alpha\beta}(\vec{x},t)\rangle_l $, $\langle \widehat{J}^{\mu}(\vec{x},t)\rangle_l$, and $\langle \widehat{S}^{\mu\alpha\beta}(\vec{x},t)\rangle_l$ can be considered as the local equilibrium contribution of the macroscopic currents appearing in the hydrodynamic description. Similarly, out of equilibrium effects encoded in $\delta\langle \widehat{T}^{\alpha\beta}(\vec{x},t)\rangle $, $\delta\langle \widehat{J}^{\mu}(\vec{x},t)\rangle$, and $\delta\langle \widehat{S}^{\mu\alpha\beta}(\vec{x},t)\rangle$ can be identified with the away from local equilibrium contribution of the macroscopic currents with the hydrodynamic framework not far away from equilibrium. This identification of $T^{\mu\nu}_{(1)}$, $J^{\mu}_{(1)}$, and $S^{\mu\alpha\beta}_{(1)}$ with the out of equilibrium contribution of the corresponding microscopic current, i.e., $\delta\langle \widehat{T}^{\alpha\beta}(\vec{x},t)\rangle $, $\delta\langle \widehat{J}^{\mu}(\vec{x},t)\rangle$, and $\delta\langle \widehat{S}^{\mu\alpha\beta}(\vec{x},t)\rangle$, respectively, allows us to write down the Kubo relations of various transport coefficients that arise in the dissipative hydrodynamic theory.  The Kubo relations can be obtained once we find $\widehat{C}^{}(\vec{x}^{\prime},t^{\prime})$ in terms of the irreducible tensor structure of $\widehat{T}^{\mu\nu}$, $\widehat{J}^{\mu}$, and $\widehat{S}^{\lambda\mu\nu}$. In the Kubo framework, it is considered that the irreducible tensor decomposition of microscopic currents is the same as the decomposition of its macroscopic counterpart. In the present manuscript, we argued two decompositions of the dissipative part of the spin tensor given in Eq.~\eqref{equ18ver1}, and \eqref{equ54new}. These two decompositions bring different sets of spin transport coefficients in the hydrodynamic description. Note that in these two decomposition frameworks, the decomposition of $T^{\mu\nu}$ and $J^{\mu}$ is the same. The difference only appears due to the decomposition of $S^{\mu\alpha\beta}_{(1)}$. Naturally, for these two decomposition schemes, the expression of $\widehat{C}(\vec{x},t)$ is different. We first obtain the expression of $\widehat{C}(\vec{x},t)$ for these decomposition schemes. First, we consider the decomposition given in Eqs.~\eqref{equ17ver1}, \eqref{equ18ver1} to get $\widehat{C}(\vec{x},t)$ and the Kubo relation of the associated transport coefficients in Sec.~\ref{kubo1}. Subsequently, in Sec.~\ref{kubo2}, we write down the Kubo relation for the transport coefficients appearing in Eq.~\eqref{equ54new}. 

\subsection{Kubo relations for the transport coefficients I}
\label{kubo1}
In this subsection, we show the Kubo relations for the transport coefficients that appear in Eqs.~\eqref{equ20ver1}-\eqref{equ27ver1}. The expression of $\widehat{C}$ depends on the tensor decompositions of various dissipative currents that appear in the $\widehat{T}^{\mu\nu}$, and $\widehat{S}^{\lambda\mu\nu}$. First, we will consider the decomposition of these tensors that have been given in Eqs.~\eqref{equ6ver1}-\eqref{equ8ver1}, and \eqref{equ17ver1}- \eqref{equ18ver1}. 
The tensor decomposition of microscopic currents, i.e. the decomposition of $\widehat{T}^{\mu\nu}$ and $\widehat{S}^{\lambda\mu\nu}$, is considered to be on an equal footing with the decomposition of the macroscopic currents $T^{\mu\nu}$ and $S^{\lambda\mu\nu}$. The term $\widehat{C}$ contains all the thermodynamic forces, and using the spin hydrodynamic equations, it can be shown that (see Appendix.~\ref{appenE} for details)
\begin{align}
\widehat{C} & =-\widehat{P}^{\star}\beta\theta
-\widehat{\mathcal{J}}^{\mu}\nabla_{\mu}\alpha+\widehat{h}^{\mu}\frac{S^{\alpha\beta}}{\varepsilon+P}\nabla_{\mu}(\beta\omega_{\alpha\beta})+\beta \widehat{\pi}^{\mu\nu}\sigma_{\mu\nu}\nonumber\\
&~~~~ -2\widehat{\Phi} u^{\alpha}\nabla^{\beta}(\beta\omega_{\alpha\beta})-2\widehat{\tau}^{\mu\beta}_{(s)}u^{\alpha}\Delta^{\gamma\rho}_{\mu\beta}\nabla_{\gamma}(\beta\omega_{\alpha\rho})-2\widehat{\tau}^{\mu\beta}_{(a)}u^{\alpha}\Delta^{[\gamma\rho]}_{[\mu\beta]}\nabla_{\gamma}(\beta\omega_{\alpha\rho}) -\widehat{\Theta}_{\mu\alpha\beta}\Delta^{\alpha\delta}\Delta^{\beta\rho}\Delta^{\mu\gamma}\nabla_{\gamma}(\beta\omega_{\delta\rho}).
\label{equ53ver1}
\end{align}
In the above equation,
\begin{align}
& \widehat{P}^{\star}=\left(\widehat{P}-\widehat{\Pi}-\widehat{\varepsilon}\gamma+\widehat{n}\gamma^{\prime}+\widehat{S}^{\alpha\beta}\gamma_{\alpha\beta}\right),\label{bulk}\\
& \widehat{\mathcal{J}}^{\mu} = \widehat{J}^{\mu}_{(1)}-\frac{n}{\varepsilon+P}\widehat{h}^{\mu}.
\end{align}
Explicit expressions of $\gamma$, $\gamma^{\prime}$, and $\gamma^{\alpha\beta}$ are given in Eqs.~\eqref{equE9new}-\eqref{equE11new}. A notable observation in Eq.~\eqref{bulk} for the bulk pressure is the emergence of a contribution from the spin susceptibility term \( \gamma_{\alpha\beta} \)~\cite{Hattori:2019ahi, Fukushima:2020ucl, Daher:2022wzf, Daher:2022xon, Biswas:2022bht}. Now one can calculate the correlation function as shown in Eq.~\eqref{delT}, \eqref{delJ} and \eqref{delS}. Here, we do not consider any anisotropy in the system created due to the spin polarization. Note that $\omega^{\mu\nu}$ is a two-rank antisymmetric tensor. Hence, it can be decomposed into electric and magnetic field-like components. In principle, these vector components can give rise to an anisotropy in the system, e.g., a strong background magnetic field produces an anisotropic structure in the electrical conductivity tensor. But here, we do not consider such strong field effects. Moreover, using Curie's principle, one can argue that the correlation function between operators of different ranks and spatial parity vanishes~\cite{Landau_Physical_kinetics}.

We emphasize that $\widehat{C}(\vec{x},t)$ contains the dissipative contribution of every conserved current. Through the Eqs.~\eqref{delT}, \eqref{delJ}, the expectation of $\widehat{P}^{\star}(\vec{x},t)$, $\widehat{\pi}^{\mu\nu}(\vec{x},t)$, and $\widehat{\mathcal{J}}^{\mu}(\vec{x},t)$ concerning non-equilibrium states gives the constitutive relation of the corresponding macroscopic currents. One can identify the known result of the dissipative term  of the stress-energy, like the bulk and the shear part of the energy-momentum tensor:
\begin{align}
    \Pi=-\langle \widehat{P}^{\star}(\vec{x},t)\rangle& =-\int d^3x^{\prime}\int_{-\infty}^t dt^{\prime}~e^{\epsilon(t^{\prime}-t)}\bigg(\widehat{P}^{\star}(\vec{x},t),\widehat{C}^{}(\vec{x}^{\prime},t^{\prime})\bigg)_l\nonumber\\
    & = \beta\theta\int d^3x^{\prime}\int_{-\infty}^t dt^{\prime}~e^{\epsilon(t^{\prime}-t)}\bigg(\widehat{P}^{\star}(\vec{x},t),\widehat{P}^{\star}(\vec{x}^{\prime},t^{\prime})\bigg)_l\label{Pid}\\
    \pi^{\mu\nu}(\vec{x},t)=\langle\widehat{\pi}^{\mu\nu}(\vec{x},t)\rangle& = \int d^3x^{\prime}\int_{-\infty}^t dt^{\prime}~e^{\epsilon(t^{\prime}-t)}\bigg(\widehat{\pi}^{\mu\nu}(\vec{x},t),\widehat{C}^{}(\vec{x}^{\prime},t^{\prime})\bigg)_l\nonumber\\
    & = \int d^3x^{\prime}\int_{-\infty}^t dt^{\prime}~e^{\epsilon(t^{\prime}-t)}\bigg(\widehat{\pi}^{\mu\nu}(\vec{x},t),\widehat{\pi}^{\rho\delta}(\vec{x}^{\prime},t^{\prime})\bigg)_l\beta\sigma_{\rho\delta}.\label{sigd}
\end{align}

The corresponding dissipative coefficients, i.e., $\zeta,\eta$ to bulk and shear dissipation, respectively, can be expressed as~\cite{Harutyunyan:2017lrm,Huang:2011dc}.
\begin{align}
    \eta&=\frac{\beta}{10}\int d^3x^{\prime}\int_{-\infty}^t dt^{\prime}~e^{\epsilon(t^{\prime}-t)}\Big(\widehat{\pi}^{\mu\nu}(x),\widehat{\pi}_{\mu\nu}(x^{\prime})\Big)_l,\\
    \zeta&=\beta\int d^3x^{\prime}\int_{-\infty}^t dt^{\prime}~e^{\epsilon(t^{\prime}-t)}\Big(\widehat{P}^{\star}(x),\widehat{P}^{\star}(x^{\prime})\Big)_l\,.
\end{align}
It is important to note that the $(0, i)$-th component of the energy-momentum tensor
gives rise to the orbital angular momentum. The $(0, i)$-th component of the energy-momentum tensor, i.e, $\widehat{h}^{\mu}$ gets coupled to the equilibrium spin density and leads to transport which breaks the degeneracy of charge conductivity and thermal conductivity. This coupling of $\widehat{h}^{\mu}$ and $S^{\alpha\beta}$ gives rise to cross conductivity between charge current and heat (a net momentum flow) current:
\begin{align}
    \mathcal{J}^{\mu}(\vec{x},t)=\langle\widehat{\mathcal{J}}^{\mu}(\vec{x},t)\rangle & = \int d^3x^{\prime}\int_{-\infty}^t dt^{\prime}~e^{\epsilon(t^{\prime}-t)}\bigg(\widehat{\mathcal{J}}^{\mu}(\vec{x},t),\widehat{C}^{}(\vec{x}^{\prime},t^{\prime})\bigg)_l\nonumber\\
    & = -\int d^3x^{\prime}\int_{-\infty}^t dt^{\prime}~e^{\epsilon(t^{\prime}-t)}\bigg(\widehat{\mathcal{J}}^{\mu}(\vec{x},t),\widehat{\mathcal{J}}^{\rho}(\vec{x}^{\prime},t^{\prime})\bigg)_l\nabla_{\rho}\alpha\nonumber\\
    &\,\,+\int d^3x^{\prime}\int_{-\infty}^t dt^{\prime}~e^{\epsilon(t^{\prime}-t)}\bigg(\widehat{\mathcal{J}}^{\mu}(\vec{x},t),\widehat{h}^{\rho}(\vec{x}^{\prime},t^{\prime})\bigg)_l\Big(\frac{S^{\alpha\beta}}{\varepsilon+P}\Big)\nabla_{\rho}(\beta\omega_{\alpha\beta}).
    \label{Jid}
\end{align}
\begin{align}
    h^{\mu}(\vec{x},t)=\langle\widehat{h}^{\mu}(\vec{x},t)\rangle & = \int d^3x^{\prime}\int_{-\infty}^t dt^{\prime}~e^{\epsilon(t^{\prime}-t)}\bigg(\widehat{h}^{\mu}(\vec{x},t),\widehat{C}^{}(\vec{x}^{\prime},t^{\prime})\bigg)_l\nonumber\\
    & = -\int d^3x^{\prime}\int_{-\infty}^t dt^{\prime}~e^{\epsilon(t^{\prime}-t)}\bigg(\widehat{h}^{\mu}(\vec{x},t),\widehat{\mathcal{J}}^{\rho}(\vec{x}^{\prime},t^{\prime})\bigg)_l\nabla_{\rho}\alpha\nonumber\\
    &\,\,+\int d^3x^{\prime}\int_{-\infty}^t dt^{\prime}~e^{\epsilon(t^{\prime}-t)}\bigg(\widehat{h}^{\mu}(\vec{x},t),\widehat{h}^{\rho}(\vec{x}^{\prime},t^{\prime})\bigg)_l\Big(\frac{S^{\alpha\beta}}{\varepsilon+P}\Big)\nabla_{\rho}(\beta\omega_{\alpha\beta}).
    \label{hid}
\end{align}
Using the last expressions, one can connect the charge and thermal transport coefficients to the correlation functions in the following way~\cite{Huang:2011dc}
\begin{align}
    \widetilde{\kappa}_{11}=&-\frac{1}{3}\int d^3x^{\prime}\int_{-\infty}^t dt^{\prime}~e^{\epsilon(t^{\prime}-t)}\bigg(\widehat{\mathcal{J}}^{\mu}(\vec{x},t),\widehat{\mathcal{J}}_{\mu}(\vec{x}^{\prime},t^{\prime})\bigg)_l\,\nonumber\\
    &=\frac{1}{3\beta}\left. \frac{\partial}{\partial \omega} \operatorname{Im} G_{\widehat{\mathcal{J}}^{\mu}, \widehat{\mathcal{J}}_{\mu}}^{R}(\mathbf{0}, \omega)\right|_{\omega \rightarrow 0},
\end{align}
\begin{align}
    \widetilde{\kappa}_{12}=\kappa_{12}=&\frac{1}{3}\int d^3x^{\prime}\int_{-\infty}^t dt^{\prime}~e^{\epsilon(t^{\prime}-t)}\bigg(\widehat{\mathcal{J}}^{\mu}(\vec{x},t),\widehat{h}_{\mu}(\vec{x}^{\prime},t^{\prime})\bigg)_l\,\nonumber\\
    &=-\frac{1}{3\beta}\left. \frac{\partial}{\partial \omega} \operatorname{Im} G_{\widehat{\mathcal{J}}^{\mu}, \widehat{h}_{\mu}}^{R}(\mathbf{0}, \omega)\right|_{\omega \rightarrow 0},
\end{align}
\begin{align}
    \kappa_{11}=&-\frac{1}{3}\int d^3x^{\prime}\int_{-\infty}^t dt^{\prime}~e^{\epsilon(t^{\prime}-t)}\bigg(\widehat{h}^{\mu}(\vec{x},t),\widehat{h}_{\mu}(\vec{x}^{\prime},t^{\prime})\bigg)_l\,,\nonumber\\
    &=\frac{1}{3\beta}\left. \frac{\partial}{\partial \omega} \operatorname{Im} G_{\widehat{h}^{\mu}, \widehat{h}_{\mu}}^{R}(\mathbf{0}, \omega)\right|_{\omega \rightarrow 0}
\end{align}
The crossed transport coefficients for thermal and charge conductivity turn out to be the same from the Kubo approach as well as in the macroscopic approach as explained in Sec-\ref{entropycurrentapp}. This can be justified because correlation functions, being a function of relative coordinates $(t-t^{\prime},|\vec{x}-\vec{x}^{\prime}|)$, make a symmetric function when the position of the operators is swapped; consistent with the Onsager conditions.
The presence of these cross modes can be seen in the macroscopic counterpart as given in Eqs.~\eqref{equ21ver1}, and \eqref{equ23ver1}. Turning our interest to finding the correlation function relates to spin transportation, which can be written down through the Kubo relations,
\begin{align}
    S_{(1)}^{\mu\alpha\beta}(\vec{x},t)=\left\langle\widehat{S}_{(1)}^{\mu\alpha\beta}(\vec{x},t) \right\rangle &= \int d^3x^{\prime}\int_{-\infty}^t dt^{\prime}~e^{\epsilon(t^{\prime}-t)}\bigg(\widehat{S}_{(1)}^{\mu\alpha\beta}(\vec{x},t),\widehat{C}^{}(\vec{x}^{\prime},t^{\prime})\bigg)_l\nonumber\\
    & = -\int d^3x^{\prime}\int_{-\infty}^t dt^{\prime}~e^{\epsilon(t^{\prime}-t)}\bigg(\widehat{S}_{(1)}^{\mu\alpha\beta}(\vec{x},t),\widehat{S}_{(1)}^{\rho\gamma\delta}(\vec{x}^{\prime},t^{\prime})\bigg)_l\nabla_{\rho}(\beta\omega_{\gamma\delta})\label{Spin1}.
\end{align}
where averages of these dissipative spin operators match with the macroscopic/hydrodynamical counterpart as
    \begin{align}
        &\Phi=\langle\widehat{\Phi}\rangle= -2\int d^3x^{\prime}\int_{-\infty}^t dt^{\prime}~e^{\epsilon(t^{\prime}-t)}\bigg(\widehat{\Phi}(\vec{x},t),\widehat{\Phi}(\vec{x}^{\prime},t^{\prime})\bigg)_l\nabla^{\delta}(\beta\omega_{\gamma\delta})u^{\gamma},\label{operatorcorre1}\\
        &\tau^{\mu\alpha}_{(s)}=\left\langle\widehat{\tau}_{(s)}^{\mu\alpha}\right\rangle=-2\int d^3x^{\prime}\int_{-\infty}^t dt^{\prime}~e^{\epsilon(t^{\prime}-t)}\bigg(\widehat{\tau}_{(s)}^{\mu\alpha}(\vec{x},t),\widehat{\tau}_{(s)}^{\rho\delta}(\vec{x}^{\prime},t^{\prime})\bigg)_l\nabla_{\rho}(\beta\omega_{\gamma\delta})u^{\gamma},\label{operatorcorre2}\\
        &\tau^{\mu\alpha}_{(a)}=\left\langle\widehat{\tau}_{(a)}^{\mu\alpha}\right\rangle=-2\int d^3x^{\prime}\int_{-\infty}^t dt^{\prime}~e^{\epsilon(t^{\prime}-t)}\bigg(\widehat{\tau}_{(a)}^{\mu\alpha}(\vec{x},t),\widehat{\tau}_{(a)}^{\rho\delta}(\vec{x}^{\prime},t^{\prime})\bigg)_l\nabla_{\rho}(\beta\omega_{\gamma\delta})u^{\gamma},\label{operatorcorre3}\\
       & \Theta^{\mu\alpha\beta}= \langle\widehat{\Theta}^{\mu\alpha\beta}\rangle=-\int d^3x^{\prime}\int_{-\infty}^t dt^{\prime}~e^{\epsilon(t^{\prime}-t)}\bigg(\widehat{\Theta}^{\mu\alpha\beta}(\vec{x},t),\widehat{\Theta}^{\rho\gamma\delta}(\vec{x}^{\prime},t^{\prime})\bigg)_l\nabla_{\rho}(\beta\omega_{\gamma\delta})\,.
       \label{operatorcorre4}
    \end{align}

The operator correlation functions appearing in Eqs.~\eqref{operatorcorre1}-\eqref{operatorcorre4} satisfy the following properties,
\begin{align}
    \bigg(\widehat{\tau}_{(s)}^{\mu\alpha}(\vec{x},t),\widehat{\tau}_{(s)}^{\rho\delta}(\vec{x}^{\prime},t^{\prime})\bigg)_l=\frac{1}{5}\Delta^{\mu\alpha,\rho\delta}\bigg(\widehat{\tau}_{(s)}^{\lambda\nu}(\vec{x},t),\widehat{\tau}_{(s)\,\lambda\nu}(\vec{x}^{\prime},t^{\prime})\bigg)_l\,,
\end{align}
\begin{align}
    \bigg(\widehat{\tau}_{(a)}^{\mu\alpha}(\vec{x},t),\widehat{\tau}_{(a)}^{\rho\delta}(\vec{x}^{\prime},t^{\prime})\bigg)_l=\frac{1}{3}\Delta^{[\mu\alpha][\rho\delta]}\bigg(\widehat{\tau}_{(a)}^{\lambda\nu}(\vec{x},t),\widehat{\tau}_{(a)\,\lambda\nu}(\vec{x}^{\prime},t^{\prime})\bigg)_l\,,
\end{align}
\begin{align}
    \bigg(\widehat{\Theta}^{\mu\alpha\beta}(\vec{x},t),\widehat{\Theta}^{\nu\gamma\delta}(\vec{x}^{\prime},t^{\prime})\bigg)_l=\frac{1}{9}\Delta^{\mu\nu}\Delta^{[\alpha\beta][\gamma\delta]}\bigg(\widehat{\Theta}^{\lambda\eta\zeta}(\vec{x},t),\widehat{\Theta}_{\lambda\eta\zeta}(\vec{x},t)\bigg)_l\,.
\end{align}
Once we have expressed the macroscopic currents appearing in the dissipative part of the spin tensor in terms of the microscopic current correlators, we can now write down the Kubo relation for the spin transport coefficients,

\begin{align}
    \chi_{1}&=\int d^3x^{\prime}\int_{-\infty}^t dt^{\prime}~e^{\epsilon(t^{\prime}-t)}\bigg(\widehat{\Phi}(\vec{x},t),\widehat{\Phi}(\vec{x}^{\prime},t^{\prime})\bigg)_l\,\nonumber\\
    &=-\frac{1}{\beta}\left. \frac{\partial}{\partial \omega} \operatorname{Im} G_{\widehat{\Phi}, \widehat{\Phi}}^{R}(\mathbf{0}, \omega)\right|_{\omega \rightarrow 0},\label{chi1Kubo}
\end{align}

\begin{align}
    \chi_{2}&=\frac{1}{5}\int d^3x^{\prime}\int_{-\infty}^t dt^{\prime}~e^{\epsilon(t^{\prime}-t)}\bigg(\widehat{\tau}_{(s)}^{\lambda\nu}(\vec{x},t),\widehat{\tau}_{(s)\,\lambda\nu}(\vec{x}^{\prime},t^{\prime})\bigg)_l\,\nonumber\\
    &=-\frac{1}{5\beta}\frac{\partial}{\partial \omega} \operatorname{Im} G_{\widehat{\tau}_{(s)}^{\mu\lambda}, \widehat{\tau}_{(s)\,\mu\lambda}}^{R}(\mathbf{0}, \omega)\Big|_{\omega \rightarrow 0},\label{chi2Kubo}
\end{align}
\begin{align}
    \chi_{3}&=\frac{1}{3}\int d^3x^{\prime}\int_{-\infty}^t dt^{\prime}~e^{\epsilon(t^{\prime}-t)}\bigg(\widehat{\tau}_{(a)}^{\lambda\nu}(\vec{x},t),\widehat{\tau}_{(a)\,\lambda\nu}(\vec{x}^{\prime},t^{\prime})\bigg)_l\,\nonumber\\
    &=-\frac{1}{3\beta}\frac{\partial}{\partial \omega} \operatorname{Im} G_{\widehat{\tau}_{(a)}^{\mu\lambda}, \widehat{\tau}_{(a)\,\mu\lambda}}^{R}(\mathbf{0}, \omega)\Big|_{\omega \rightarrow 0},\label{chi3Kubo}
\end{align}

\begin{align}
    \chi_{4}&=-\frac{1}{9}\int d^3x^{\prime}\int_{-\infty}^t dt^{\prime}~e^{\epsilon(t^{\prime}-t)}\bigg(\widehat{\Theta}^{\lambda\eta\zeta}(\vec{x},t),\widehat{\Theta}_{\lambda\eta\zeta}(\vec{x},t)\bigg)_l\,\nonumber\\
    &=\frac{1}{9\beta}\frac{\partial}{\partial \omega} \operatorname{Im} G_{\widehat{\Theta}_{\lambda\eta\zeta},\widehat{\Theta}^{\lambda\eta\zeta}}^{R}(\mathbf{0}, \omega)\Big|_{\omega \rightarrow 0}\,.\label{chi4Kubo}
\end{align}

Constitutive relations for the dissipative part of the energy-momentum tensor, charged currents, and the spin tensor, along with the respective transport coefficient, completely specify the spin hydrodynamic theory. Using the Kubo relations, one can compute the transport coefficients once we specify the microscopic interactions among the constituents.

In this article, we introduced two different approaches to decomposing the dissipative part of the spin tensor.  According to this alternative decomposition, as given in Eq.~\eqref{equ54new}, different sets of transport coefficients appear in the theory. In the next section, we find the Kubo relation for the alternative tensor decomposition. Moreover, we also connect the spin transport coefficients arising in these two frameworks. 

\subsection{Kubo relations for the transport coefficients II}
\label{kubo2}
We start with the general structure for the spin operator, $\widehat{S}_{(1)}^{\mu\alpha\beta}$ that gives rise to the dissipative part in the macroscopic description \footnote{It is important to note that the decomposition of the energy-momentum tensor in a relativistic theory is guided by the decomposition of its non-relativistic counterpart, where we can write the dissipative part of the energy-momentum tensor in terms of the bulk and the shear terms. But in spin transport, we do not have any prior guiding principle for the decomposition of the spin tensor, except the orthogonality condition of the dissipative spin current operator, i.e., $u_{\mu}\widehat{S}^{\mu\alpha\beta}_{(1)}=0$. This helps us to generically write $\widehat{S}^{\mu\alpha\beta}_{(1)}$ as given in Eq.~\eqref{decomS}, where $\widehat{\Xi}^{\mu\alpha\beta}$ and $\widehat{\mathcal{V}}^{\mu[\alpha}u^{\beta]}$ spans the $(i,[j ,k])$ and $(i,[j,0])$ components of the dissipative current in rest frame $u^{\mu}_{0}=(1,\vec{0})$ respectively. } 
\begin{align}
    \widehat{S}_{(1)}^{\mu\alpha\beta}=\widehat{\Xi}^{\mu\alpha\beta}+\widehat{\mathcal{V}}^{\mu[\alpha}u^{\beta]} \label{decomS}\,,
\end{align}
Two different parts of the $\widehat{S}_{(1)}^{\mu\alpha\beta}$ satisfy the condition $u_{\mu}\widehat{\Xi}^{\mu\alpha\beta}=0, u_{\alpha}\widehat{\Xi}^{\mu\alpha\beta}=0,\,u_{\mu}\widehat{\mathcal{V}}^{\mu\alpha}=0, \widehat{\mathcal{V}}^{\mu\alpha}u_{\alpha}=0$. Interestingly using the general decomposition of the spin tensor Eq.~\eqref{decomS} one can get a generic expression the operator $\widehat{C}$,
\begin{align}
    \widehat{C} & =-\widehat{P}^{\star}\beta\theta
-\widehat{\mathcal{J}}^{\mu}\nabla_{\mu}\alpha+\widehat{h}^{\mu}\frac{S^{\alpha\beta}}{\varepsilon+P}\nabla_{\mu}(\beta\omega_{\alpha\beta})+\beta \widehat{\pi}^{\mu\nu}\sigma_{\mu\nu}-\widehat{\Xi}^{\mu\alpha\beta}\nabla_{\mu}(\beta\omega_{\alpha\beta})-\widehat{\mathcal{V}}^{\mu\alpha}u^{\beta}\nabla_{\mu}(\beta\omega_{\alpha\beta}).
\label{generalC}
\end{align}
The first four terms of the above equation are similar to the respective terms that arise  Eq.~\eqref{equ53ver1}. These terms 
leads to the flow of momentum, heat, and charge as depicted in Eqs.~\eqref{Pid}-\eqref{hid}.
The last two terms in Eq.~\eqref{generalC} manifest the spin transport in a different way as compared to Eq.~\eqref{equ53ver1}. We emphasize that all possible dissipative currents and the corresponding transport coefficients can also be extracted using Eq.~\eqref{generalC}. 
We can write down the Kubo relations for the spin transport coefficients using the following expression, 
\begin{align}
    S_{(1)}^{\mu\alpha\beta}&=\left\langle\widehat{S}_{(1)}^{\mu\alpha\beta}\right\rangle = -\int d^3x^{\prime}\int_{-\infty}^t dt^{\prime}~e^{\epsilon(t^{\prime}-t)}\bigg(\widehat{S}_{(1)}^{\mu\alpha\beta}(\vec{x},t),\widehat{S}_{(1)}^{\rho\gamma\delta}(\vec{x}^{\prime},t^{\prime})\bigg)_l\nabla_{\rho}(\beta\omega_{\gamma\delta}).
    \label{S1general}
\end{align}
On the other hand, by comparing Eq.~\eqref{S1general} with Eq.~\eqref{eq:dcom_d_SAM} we can identify the six rank tensor $\Sigma^{\mu\alpha\beta\rho\gamma\delta}$ that contains all the dissipative currents and the associated transport coefficients,
\begin{align}
    \Sigma^{\mu\alpha\beta\rho\gamma\delta}& = -\int d^3x^{\prime}\int_{-\infty}^t dt^{\prime}~e^{\epsilon(t^{\prime}-t)}\bigg(\widehat{S}_{(1)}^{\mu\alpha\beta}(\vec{x},t),\widehat{S}_{(1)}^{\rho\gamma\delta}(\vec{x}^{\prime},t^{\prime})\bigg)_l\,.
    \label{equ101new}
\end{align}
It is important to note that, we can isolate $\widehat{\mathcal{V}}^{\mu \alpha}$, and  $\widehat{\Xi}^{\mu\alpha\beta}$ from Eq.~\eqref{decomS} by taking the projection of $\widehat{S}^{\mu\alpha\beta}_{(1)}$ along $u^{\beta}$, and orthogonal to  $u^{\beta}$, respectively. Since the microscopic operators and their macroscopic counterparts have the same tensor decomposition, we can obtain the expressions for the macroscopic counterpart of these tensors, i.e., $\left\langle\widehat{\mathcal{V}}^{\mu \alpha}\right\rangle$ and $\langle\widehat{\Xi}^{\mu\alpha\beta}\rangle$ by taking the appropriate projection of $S_{(1)}^{\mu\alpha\beta}$ as given in Eq.~\eqref{equ18ver1}, and Eq.~\eqref{equ54new}. If we use the expression of $S_{(1)}^{\mu\alpha\beta}$ given in Eq.~\eqref{equ18ver1} then one obtains, 
\begin{align}
\left\langle\widehat{\mathcal{V}}^{\mu \alpha}\right\rangle = -\left(\Delta^{\mu\alpha}\Phi+\tau^{\mu\alpha}_{(s)}+\tau^{\mu\alpha}_{(a)}\right)\,.
\label{equ102new}
\end{align}
Similarly, if we take the appropriate projection of the macroscopic counterpart of $\widehat{S}_{(1)}^{\mu\alpha\beta}$ as defined alternatively in Eq.~\eqref{equ54new}, then we find,  
\begin{align}
\left\langle\widehat{\mathcal{V}}^{\mu \alpha}\right\rangle = \Lambda_{1}\Delta^{\mu\alpha}\Phi_{||}+\Lambda_{2}\Gamma_{s}^{\mu\alpha}+\Lambda_{3}\Gamma_{a}^{\mu\alpha}.
\label{equ103new}
\end{align}
In a similar way, taking the orthogonal projection (orthogonal to fluid flow) of $S_{(1)}^{\mu\alpha\beta}$ as defined in Eq.~\eqref{equ18ver1}, we find, 
\begin{align}
\langle\widehat{\Xi}^{\mu\alpha\beta}\rangle = \Theta^{\mu\alpha\beta}. 
\label{equ104new}
\end{align}
Using the alternative definition of $S_{(1)}^{\mu\alpha\beta}$ given in Eq.~\eqref{equ54new} and taking its orthogonal projection with respect to $u^{\beta}$ we find, 
\begin{align}
\langle\widehat{\Xi}^{\mu\alpha\beta}\rangle&=-\big(\Sigma_{1}+\Sigma_{3}\big)\Delta^{\mu[\alpha}\Phi^{\beta]}_{\perp}+\big(\Sigma_{2}+\Sigma_{3}\big)\varepsilon^{\mu\alpha\beta}\varphi+\Sigma_{3}\Phi^{\mu\alpha\beta}
\label{equ105new}
\end{align}
From Eqs.~\eqref{equ102new}-\eqref{equ105new}, we can conclude that Eq.~\eqref{decomS} can be used to obtain two different tensor decomposition of $S^{\mu\alpha\beta}_{(1)}$ and its microscopic counterpart.

The Kubo relation for the spin transport coefficients appearing in Eq.~\eqref{equ103new}, and Eq.~\eqref{equ105new} can be extracted from Eq.~\eqref{equ101new} by multiplying projection tensor $\mathcal{P}^{\prime}$ and $\mathcal{Q}^{\prime}$, 
\begin{align}
    \Sigma_{1}&=\frac{1}{3}\int d^3x^{\prime}\int_{-\infty}^t dt^{\prime}~e^{\epsilon(t^{\prime}-t)}\bigg(\widehat{\Xi}^{~\rho \sigma}_{\rho}(\vec{x},t),\widehat{\Xi}_{~\lambda\sigma}^{\lambda}(\vec{x}^{\prime},t^{\prime})\bigg)_l\,\nonumber\,
    \\
    &=-\left.\frac{1}{3\beta} \frac{\partial}{\partial \omega} \operatorname{Im}G_{ \widehat{\Xi}^{~\rho \sigma}_{\rho}, \widehat{\Xi}_{~\lambda\sigma}^{\lambda}}^{R}(\mathbf{0}, \omega)\right|_{\omega \rightarrow 0},
\end{align}

\begin{align}
    \Sigma_{2}&=-\frac{1}{6}\int d^3x^{\prime}\int_{-\infty}^t dt^{\prime}~e^{\epsilon(t^{\prime}-t)}\bigg(\varepsilon^{\mu\alpha\beta}\widehat{\Xi}_{\mu\alpha\beta}(\vec{x},t),\varepsilon^{\eta\gamma\delta}\widehat{\Xi}_{\eta\gamma\delta}(\vec{x}^{\prime},t^{\prime})\bigg)_l\,\nonumber\,,\\
    &=\frac{1}{6\beta}\frac{\partial}{\partial \omega} \operatorname{Im} G_{\varepsilon^{\mu\alpha\beta}\widehat{\Xi}_{\mu \alpha\beta}, \epsilon^{\eta\gamma\delta}\widehat{\Xi}_{\eta \gamma\delta}}^{R}(\mathbf{0}, \omega)\Big|_{\omega \rightarrow 0} 
\end{align}

\begin{align}
    \Sigma_{3}&=-\frac{1}{5}\int d^3x^{\prime}\int_{-\infty}^t dt^{\prime}~e^{\epsilon(t^{\prime}-t)}\bigg(\widehat{\Xi}_{\mu\alpha\beta}(\vec{x},t),\widehat{\Xi}^{\mu\alpha\beta}(\vec{x}^{\prime},t^{\prime})\bigg)_l+\frac{3\Sigma_{1}}{5}+\frac{\Sigma_{2}}{5}\,,\nonumber\\
    &=\frac{1}{5\beta}\frac{\partial}{\partial \omega} \operatorname{Im} G_{\widehat{\Xi}_{\mu \alpha\beta},\widehat{\Xi}^{\mu\alpha\beta}}^{R}(\mathbf{0}, \omega)\Big|_{\omega \rightarrow 0}+\frac{3\Sigma_{1}}{5}+\frac{\Sigma_{2}}{5}\,,
\end{align}

\begin{align}
    \Lambda_{1}&=-\frac{4}{9}\int d^3x^{\prime}\int_{-\infty}^t dt^{\prime}~e^{\epsilon(t^{\prime}-t)}\bigg(\widehat{\mathcal{V}}^{\mu}_{~\mu}(\vec{x},t),\widehat{\mathcal{V}}^{\lambda}_{~\lambda}(\vec{x}^{\prime},t^{\prime})\bigg)_l\nonumber\,,\\
    &=\left.\frac{4}{9\beta} \frac{\partial}{\partial \omega} \operatorname{Im} G_{\widehat{\mathcal{V}}^{\mu}_{~\mu}, \widehat{\mathcal{V}}^{\lambda}_{~\lambda}}^{R}(\mathbf{0}, \omega)\right|_{\omega \rightarrow 0},
\end{align}

\begin{align}
    \Lambda_{2}&=-\frac{2}{5}\int d^3x^{\prime}\int_{-\infty}^t dt^{\prime}~e^{\epsilon(t^{\prime}-t)}\bigg(\widehat{\mathcal{V}}^{\mu\lambda}_{(s)}(\vec{x},t),\widehat{\mathcal{V}}_{(s)~\mu\lambda}(\vec{x}^{\prime},t^{\prime})\bigg)_l\nonumber\,,\\
    &=\frac{2}{5\beta} \frac{\partial}{\partial \omega} \operatorname{Im} G_{\widehat{\mathcal{V}}_{(s)}^{\mu\lambda}, \widehat{\mathcal{V}}_{(s)~\mu\lambda}}^{R}(\mathbf{0}, \omega)\Big|_{\omega \rightarrow 0},
\end{align}

\begin{align}
   \Lambda_{3}&=-\frac{2}{3}\int d^3x^{\prime}\int_{-\infty}^t dt^{\prime}~e^{\epsilon(t^{\prime}-t)}\bigg(\widehat{\mathcal{V}}^{\mu\lambda}_{(a)}(\vec{x},t),\widehat{\mathcal{V}}_{(a)~\mu\lambda}(\vec{x}^{\prime},t^{\prime})\bigg)_l\nonumber\,,\\
   &=\frac{2}{3\beta}\frac{\partial}{\partial \omega} \operatorname{Im} G_{\widehat{\mathcal{V}}_{(a)}^{\mu\lambda}, \widehat{\mathcal{V}}_{(a)\,\mu\lambda}}^{R}(\mathbf{0}, \omega)\Big|_{\omega \rightarrow 0}\,,
\end{align}
here, $\mathcal{V}^{\mu\alpha}_{(s)}$ and  $\mathcal{V}^{\mu\alpha}_{(a)}$ are the trace-less symmetric and anti-symmetric part of the $\mathcal{V}^{\mu\alpha}$ respectively. In a similar way, we can write down the Kubo relations for the transport coefficients that appear in the constitutive relation of the dissipative currents $\Phi$, $\tau^{\mu\alpha}_{(s)}$, $\tau^{\mu\alpha}_{(a)}$, and $\Theta^{\mu\alpha\beta}$ (see Eq.~\eqref{equ102new}, and \eqref{equ104new}),  
\begin{align}
    \chi_{1}=-\left.\frac{1}{9\beta} \frac{\partial}{\partial \omega} \operatorname{Im} G_{\widehat{\mathcal{V}}^{\mu}_{~\mu}, \widehat{\mathcal{V}}^{\lambda}_{~\lambda}}^{R}(\mathbf{0}, \omega)\right|_{\omega \rightarrow 0},
\end{align}

\begin{align}
    \chi_{2}=-\frac{1}{5\beta} \frac{\partial}{\partial \omega} \operatorname{Im} G_{\widehat{\mathcal{V}}_{(s)}^{\mu\lambda}, \widehat{\mathcal{V}}_{(s)\,\mu\lambda}}^{R}(\mathbf{0}, \omega)\Big|_{\omega \rightarrow 0},
\end{align}

\begin{align}
    \chi_{3}=-\frac{1}{3\beta} \frac{\partial}{\partial \omega} \operatorname{Im} G_{\widehat{\mathcal{V}}_{(a)}^{\mu\lambda}, \widehat{\mathcal{V}}_{(a)\,\mu\lambda}}^{R}(\mathbf{0}, \omega)\Big|_{\omega \rightarrow 0},
\end{align}

\begin{align}
    \chi_{4}=\frac{1}{9\beta}\frac{\partial}{\partial \omega} \operatorname{Im} G_{\widehat{\Xi}_{\mu \alpha\beta},\widehat{\Xi}^{\mu\alpha\beta}}^{R}(\mathbf{0}, \omega)\Big|_{\omega \rightarrow 0}\,.
\end{align}
Interestingly by comparing the Kubo relations of both sets of transport coefficients, i.e., $\left\{\Sigma_1,\Sigma_2,\Sigma_3,\Lambda_1, \Lambda_2,\Lambda_3 \right\}$ and $\left\{\chi_1,\chi_2,\chi_3,\chi_4\right\}$ we observe that 
\begin{align}
\Lambda_{1}=-4\chi_{1}\,,\Lambda_{2}=-2\chi_{2}\,,\Lambda_{3}=-2\chi_{3}\,,   \chi_{4}=\frac{5}{9}\Sigma_{3}-\frac{3}{9}\Sigma_{1}-\frac{1}{9}\Sigma_{2}\,.
\end{align}
The expression of $\chi_4$ written in terms of $\Sigma_{1}$, $\Sigma_{2}$, and $\Sigma_{3}$, reflects further reducible tensor structure of $\Theta^{\mu\alpha\beta}$. We emphasize that the dissipative current associated with $\chi_{4}$ has nine degrees of freedom. On the other hand, the dissipative currents associated with $\Sigma_{1}$, $\Sigma_{2}$, and $\Sigma_{3}$, represent three, one, and five degrees of freedom, respectively. The above relation of $\chi_{4}$ in terms of $\Sigma_{1}$, $\Sigma_{2}$, and $\Sigma_{3}$ is the manifestation of the matching of degrees of freedom in two different tensor decomposition approaches.

\section{Summary and outlook}
\label{sec4}

In this article, we explored a first-order dissipative spin hydrodynamic framework, where the spin chemical potential is treated as the leading order term in the hydrodynamic gradient expansion, i.e., $\omega^{\mu\nu} \sim \mathcal{O}(1)$. The entropy current analysis reveals that, within this spin hydrodynamic framework, the energy-momentum tensor must be symmetric at the order of $\mathcal{O}(\partial)$. By applying the entropy current analysis, we obtained the constitutive relations for various dissipative currents in terms of the first-order gradients of the hydrodynamic variables: temperature $(T)$, chemical potential ($\mu$), and spin chemical potential ($\omega^{\mu\nu}$). 
A key novelty emerging from our analysis is the entropy production due to the coupling between the $(0, i)$ component of the energy-momentum tensor and the spin density tensor $(S^{\alpha\beta})$. We interpret this as dissipation caused by spin-orbit coupling, where the $(0, i)$ component contributes to the orbital part of the energy-momentum tensor. This interaction introduces complexities in defining the fluid frame. Traditionally, two equivalent fluid frames are used to determine the fluid flow vector: the Landau frame and the Eckart frame. In standard hydrodynamics, these frames are equivalent. However, this equivalence is not obviously manifested due to spin-orbit interaction, as shown in our analysis. 
Detailed investigations will be necessary to look into this intriguing aspect related to the fluid frame choice in spin hydrodynamics. Our calculation also shows the presence of cross-diffusion coefficients. 
These cross-diffusion coefficients connect the gradients of both the spin chemical potential and the ordinary chemical potential. The presence of such cross-diffusion coefficients implies that the vector dissipative currents are not solely related to the gradients of the corresponding chemical potentials through a simple diagonal matrix of transport coefficients. Instead, the transport coefficient matrix contains off-diagonal elements, which generate cross-diffusion currents.

Throughout the analysis, the decomposition of the energy-momentum tensor into irreducible dissipative currents, such as bulk and shear tensor components, is unique. However, we propose two distinct methods for decomposing the dissipative part of the spin tensor. These two decompositions correspond to different forms of dissipative currents and different sets of spin transport coefficients. By employing a microscopic description of the energy-momentum tensor, charge current, and spin tensor, we derived the Kubo relations for the different transport coefficients involved in our calculations. Additionally, we demonstrate the equivalence of these tensor decompositions of the spin tensor by establishing the equivalence of the corresponding dissipative currents and their associated transport coefficients. The next step is to estimate these transport coefficients, with a particular focus on the spin transport coefficients, using realistic field theory models, e.g., QCD-inspired effective models involving interacting spin half fields. These estimates can then be incorporated into a realistic spin hydrodynamic code, which is currently in development. This will allow for a deeper exploration of the fermion spin polarization within an evolving QCD medium. The present study might have a general applicability, e.g., in the condensed matter spin systems. In recent studies, the hydrodynamic as well as spin hydrodynamic behavior of electrons has gained significant attention, as explored in works such as \cite{Lucas:2017idv, Torre:2015pva, DiSante:2019zrd, Narozhny:2022ncn,Narozhny:2019uib, Hui:2020lzz,Jaiswal:2024urq}. This opens up exciting opportunities for future extension of the present investigations, particularly in applying our formalism to extract spin transport coefficients in these condensed matter systems. In this context, a detailed comparative study between the weak-coupling regime \cite{Jaiswal:2024urq,Aung:2023vrr} and the strong-coupling limit \cite{Romatschke:2007mq, Romatschke:2021imm, Romatschke:2023ztk, Weiner:2022kgx} of the theory will be essential. Such studies would provide valuable insights, potentially revealing distinct transport behaviors in these different coupling scenarios.

\section*{Acknowledgements}
The authors acknowledge Amarersh Jaiswal and Samapan Bhadury for helpful and critical discussions regarding the content of the article. We would also like to thank the organizers of the ``Dynamics of QCD Matter (DQM) 2023" meeting, held at NISER, Bhubaneswar, India, for their warm hospitality and for providing an excellent platform that facilitated valuable discussions and advancements related to this work. A. D. acknowledges the New Faculty Seed Grant (NFSG), NFSG/PIL/2024/P3825, provided by the Birla Institute of Technology and Science Pilani, Pilani Campus, India. 
\appendix
\section{Derivation of Eq.~\eqref{equ14ver1}}
\label{appenA}
We start with the expression of $\mathcal{S}^{\mu}_{(0)}$ as given in Eq.~\eqref{equ13ver1}, 
\begin{align}
\mathcal{S}^{\mu}_{(0)} & =T^{\mu\nu}_{(0)}\beta_{\nu}+P\beta^{\mu}-\alpha J^{\mu}_{(0)}-\Omega_{\alpha\beta}S^{\mu\alpha\beta}_{(0)}.
\label{appenA1ver1}
\end{align}
The four divergence of $\mathcal{S}^{\mu}_{(0)}$ is then, 
\begin{align}
\partial_{\mu}\mathcal{S}^{\mu}_{(0)} & =(\partial_{\mu}\beta_{\nu})T^{\mu\nu}_{(0)}+\beta_{\nu}\partial_{\mu}T^{\mu\nu}_{(0)}+\beta^{\mu}\partial_{\mu}P+P\partial_{\mu}\beta^{\mu}-J^{\mu}_{(0)}\partial_{\mu}\alpha-\alpha \partial_{\mu}J^{\mu}_{(0)}-S^{\mu\alpha\beta}_{(0)}\partial_{\mu}\Omega_{\alpha\beta}-\Omega_{\alpha\beta} \partial_{\mu}S^{\mu\alpha\beta}_{(0)}\nonumber\\
& =(\partial_{\mu}\beta_{\nu})T^{\mu\nu}_{(0)}-\beta_{\nu}\partial_{\mu}T^{\mu\nu}_{(1)}+\beta^{\mu}\partial_{\mu}P+P\partial_{\mu}\beta^{\mu}-J^{\mu}_{(0)}\partial_{\mu}\alpha+\alpha \partial_{\mu}J^{\mu}_{(1)}-S^{\mu\alpha\beta}_{(0)}\partial_{\mu}\Omega_{\alpha\beta}+2\Omega_{\alpha\beta} T^{[\alpha\beta]}_{(1)}.
\label{appenA2ver1}
\end{align}
Note that $\mathcal{S}^{\mu}_{(0)}$ contains terms of the order $\mathcal{O}(1)$. Hence $\partial_{\mu}\mathcal{S}^{\mu}_{(0)}$ can contain terms up to order $\mathcal{O}(\partial)$. Therefore dropping all the $\mathcal{O}(\partial^2)$ terms from Eq.~\eqref{appenA2ver1} one finds, 
\begin{align}
\partial_{\mu}\mathcal{S}^{\mu}_{(0)} & =(\partial_{\mu}\beta_{\nu})T^{\mu\nu}_{(0)}+\beta^{\mu}\partial_{\mu}P+P\partial_{\mu}\beta^{\mu}-J^{\mu}_{(0)}\partial_{\mu}\alpha-S^{\mu\alpha\beta}_{(0)}\partial_{\mu}\Omega_{\alpha\beta}+2\Omega_{\alpha\beta} T^{[\alpha\beta]}_{(1)}\nonumber\\
& = (\partial_{\mu}\beta_{\nu})\left(\varepsilon u^{\mu}u^{\nu}-P\Delta^{\mu\nu}\right)+\beta DP+P\partial_{\mu}\beta^{\mu} -n D\alpha-S^{\alpha\beta} D\Omega_{\alpha\beta}+2\Omega_{\alpha\beta} T^{[\alpha\beta]}_{(1)}.
\label{appenA3ver1}
\end{align}
Here the notation $D\equiv u^{\mu}\partial_{\mu}$. Using the thermodynamic relations Eqs.~\eqref{equ9ver1}-\eqref{equ11ver1} in Eq.~\eqref{appenA3ver1} it is straight forward to show that, 
\begin{align}
\partial_{\mu}\mathcal{S}^{\mu}_{(0)}=2\beta\omega_{\alpha\beta}T^{[\alpha\beta]}_{(1)}.
\label{appenA4ver1}
\end{align}
Therefore starting from Eq.~\eqref{appenA1ver1} we arrive at Eq.~\eqref{equ14ver1}. 

\section{Derivation of Eq.~\eqref{equ15ver1}}
\label{appenB}
We start with the divergence of the entropy four-current,
\begin{align}
\partial_{\mu}\mathcal{S}^{\mu} & = (\partial_{\mu}\beta_{\nu})T^{\mu\nu}+\beta_{\nu}\partial_{\mu}T^{\mu\nu}+\beta^{\mu}\partial_{\mu}P+P\partial_{\mu}\beta^{\mu}-J^{\mu}\partial_{\mu}\alpha-\alpha \partial_{\mu}J^{\mu}-S^{\mu\alpha\beta}\partial_{\mu}\Omega_{\alpha\beta}-\Omega_{\alpha\beta} \partial_{\mu}S^{\mu\alpha\beta}\nonumber\\
& = (\partial_{\mu}\beta_{\nu})T^{\mu\nu}+\beta^{\mu}\partial_{\mu}P+P\partial_{\mu}\beta^{\mu}-J^{\mu}\partial_{\mu}\alpha-S^{\mu\alpha\beta}\partial_{\mu}\Omega_{\alpha\beta}+2\Omega_{\alpha\beta}T^{[\alpha\beta]}_{(1)} \nonumber\\
& = \left(\varepsilon u^{\mu}u^{\nu}-P\Delta^{\mu\nu}\right)\partial_{\mu}\beta_{\nu}+T^{\{\mu\nu\}}_{(1)}\partial_{\{\mu}\beta_{\nu\}}+T^{[\mu\nu]}_{(1)}\partial_{[\mu}\beta_{\nu]}+\beta DP+PD\beta+\beta P \theta\nonumber\\
& ~~~~-n\left(\beta D\mu+\mu D\beta\right)-J^{\mu}_{(1)}\partial_{\mu}\alpha-S^{\alpha\beta}\left(\beta D\omega_{\alpha\beta}+\omega^{\alpha\beta}D\beta\right)-S^{\mu\alpha\beta}_{(1)}\partial_{\mu}\Omega_{\alpha\beta}+2\Omega_{\alpha\beta}T^{[\alpha\beta]}_{(1)}\nonumber\\
& = \left(\varepsilon+P-n\mu-S^{\alpha\beta}\omega_{\alpha\beta}\right)D\beta+\beta DP-\beta nD\mu-\beta S^{\alpha\beta}D\omega_{\alpha\beta}\nonumber\\
& ~~~~+T^{\{\mu\nu\}}_{(1)}\partial_{\{\mu}\beta_{\nu\}}+T^{[\mu\nu]}_{(1)}\partial_{[\mu}\beta_{\nu]}-J^{\mu}_{(1)}\partial_{\mu}\alpha-S^{\mu\alpha\beta}_{(1)}\partial_{\mu}\Omega_{\alpha\beta}+2\Omega_{\alpha\beta}T^{[\alpha\beta]}_{(1)}\nonumber\\
& = T^{\{\mu\nu\}}_{(1)}\partial_{\{\mu}\beta_{\nu\}}+T^{[\mu\nu]}_{(1)}\partial_{[\mu}\beta_{\nu]}-J^{\mu}_{(1)}\partial_{\mu}\alpha-S^{\mu\alpha\beta}_{(1)}\partial_{\mu}\Omega_{\alpha\beta}+2\Omega_{\alpha\beta}T^{[\alpha\beta]}_{(1)}
\label{appenB1ver1}
\end{align}
We have used the thermodynamic relations Eqs.~\eqref{equ9ver1}-\eqref{equ11ver1} to obtain the final expression. 

\section{Derivation of Eq.~\eqref{equ35ver1}}
\label{appenC}
We start with the expression $\widehat{Z}(\vec{x},t)$ as given in Eq.~\eqref{equ30ver1},
\begin{align}
\int d^3x ~\widehat{Z}(\vec{x},t) & = \epsilon \int d^3x  \int _{-\infty}^t~dt^{\prime}~e^{\epsilon(t^{\prime}-t)}\bigg[\beta^{\nu}(\vec{x},t^{\prime})\widehat{T}_{0\nu}(\vec{x},t^{\prime})-\alpha (\vec{x},t^{\prime})\widehat{J}^0(\vec{x},t^{\prime})-\Omega_{\rho\sigma}(\vec{x},t^{\prime})\widehat{S}^{0\rho\sigma}(\vec{x},t^{\prime})\bigg]\nonumber\\
& = \int d^3x \bigg[\beta_{\nu}(\vec{x},t^{})\widehat{T}^{0\nu}(\vec{x},t^{})-\alpha (\vec{x},t^{})\widehat{J}^0(\vec{x},t^{})-\Omega_{\rho\sigma}(\vec{x},t^{})\widehat{S}^{0\rho\sigma}(\vec{x},t^{})\bigg]\nonumber\\
& ~~~-\int d^3x  \int _{-\infty}^t~dt^{\prime}~e^{\epsilon(t^{\prime}-t)}\bigg[\widehat{T}^{0\nu}(\vec{x},t^{\prime})\partial_{t^{\prime}}\beta_{\nu}(\vec{x},t^{\prime})+\beta_{\nu}(\vec{x},t^{\prime})\partial_{t^{\prime}}\widehat{T}^{0\nu}(\vec{x},t^{\prime})\nonumber\\
& ~~~-\widehat{J}^0(\vec{x},t^{\prime})\partial_{t^{\prime}}\alpha(\vec{x},t^{\prime})-\alpha(\vec{x},t^{\prime})\partial_{t^{\prime}}\widehat{J}^0(\vec{x},t^{\prime})
-\widehat{S}^{0\rho\sigma}(\vec{x},t^{\prime})\partial_{t^{\prime}}\Omega_{\rho\sigma}(\vec{x},t^{\prime})-\Omega_{\rho\sigma}(\vec{x},t^{\prime})\partial_{t^{\prime}}\widehat{S}^{0\rho\sigma}(\vec{x},t^{\prime})\bigg]\nonumber\\
& = \int d^3x \bigg[\beta_{\nu}(\vec{x},t^{})\widehat{T}^{0\nu}(\vec{x},t^{})-\alpha (\vec{x},t^{})\widehat{J}^0(\vec{x},t^{})-\Omega_{\rho\sigma}(\vec{x},t^{})\widehat{S}^{0\rho\sigma}(\vec{x},t^{})\bigg]\nonumber\\
& ~~~-\int d^3x  \int _{-\infty}^t~dt^{\prime}~e^{\epsilon(t^{\prime}-t)}\bigg[\widehat{T}^{0\nu}(\vec{x},t^{\prime})\partial_{t^{\prime}}\beta_{\nu}(\vec{x},t^{\prime})-\beta_{\nu}(\vec{x},t^{\prime})\partial_{i}\widehat{T}^{i\nu}(\vec{x},t^{\prime})\nonumber\\
& ~~~-\widehat{J}^0(\vec{x},t^{\prime})\partial_{t^{\prime}}\alpha(\vec{x},t^{\prime})+\alpha(\vec{x},t^{\prime})\partial_{i}\widehat{J}^i(\vec{x},t^{\prime})
-\widehat{S}^{0\rho\sigma}(\vec{x},t^{\prime})\partial_{t^{\prime}}\Omega_{\rho\sigma}(\vec{x},t^{\prime})+\Omega_{\rho\sigma}(\vec{x},t^{\prime})\partial_{i}\widehat{S}^{i\rho\sigma}(\vec{x},t^{\prime})\bigg]\nonumber\\
& = \int d^3x \bigg[\beta_{\nu}(\vec{x},t^{})\widehat{T}^{0\nu}(\vec{x},t^{})-\alpha (\vec{x},t^{})\widehat{J}^0(\vec{x},t^{})-\Omega_{\rho\sigma}(\vec{x},t^{})\widehat{S}^{0\rho\sigma}(\vec{x},t^{})\bigg]\nonumber\\
& ~~~-\int d^3x  \int _{-\infty}^t~dt^{\prime}~e^{\epsilon(t^{\prime}-t)}\bigg[\widehat{T}^{\mu\nu}(\vec{x},t^{\prime})\partial_{\mu}\beta_{\nu}(\vec{x},t^{\prime})-\widehat{J}^{\mu}(\vec{x},t^{\prime})\partial_{\mu}\alpha(\vec{x},t^{\prime})
-\widehat{S}^{\mu\rho\sigma}(\vec{x},t^{\prime})\partial_{\mu}\Omega_{\rho\sigma}(\vec{x},t^{\prime})\bigg]
\end{align}
To obtain the last line, we have dropped boundary terms and used the conservation equations, 
\begin{align}
\partial_{\mu}\widehat{T}^{\mu\nu}(\vec{x},t^{})=0; ~~\partial_{\mu}\widehat{J}^{\mu}(\vec{x},t^{})=0; ~~\partial_{\lambda}\widehat{S}^{\lambda\mu\nu}(\vec{x},t^{})=0.
\end{align}

\section{Nonequilibrium statistical operator in the linear response theory}
\label{appenD}
We start with Eq.~\eqref{equ38ver1}, i.e., 
\begin{align}
\widehat{\rho}(t)=\frac{\exp\left(-\widehat{A}+\widehat{B}\right)}{\Tr \exp\left(-\widehat{A}+\widehat{B}\right)}. 
\end{align}
In the linear response theory, we can expand the exponential of operators in the following way~\cite{alma991021547569703276}, 
\begin{align}
e^{-\widehat{A}+\widehat{B}}\simeq e^{-\widehat{A}}+ \int_0^1~d\tau ~e^{-\tau\widehat{A}} \widehat{B} e^{\tau \widehat{A}}e^{-\widehat{A}}+..... 
\end{align}
Here ``....." contains higher order terms in $\widehat{B}$ which we ignore for the linear response theory. Therefore, up to linear order  $\widehat{B}$ the density operator boils down to,
\begin{align}
\widehat{\rho}(t)& \simeq\frac{e^{-\widehat{A}}+ \int_0^1~d\tau ~e^{-\tau\widehat{A}} \widehat{B} e^{\tau \widehat{A}}e^{-\widehat{A}}}{\Tr e^{-\widehat{A}}+\Tr \bigg[\int_0^1~d\tau ~e^{-\tau\widehat{A}} \widehat{B} e^{\tau \widehat{A}}e^{-\widehat{A}}\bigg]}\nonumber\\
& \simeq \bigg(1+\int_0^1~d\tau~e^{-\tau\widehat{A}}\widehat{B}e^{\tau\widehat{A}}\bigg) 
\bigg(1-\int_0^1~d\tau~\frac{\Tr\left[e^{-\widehat{A}}e^{-\tau\widehat{A}}\widehat{B}e^{\tau\widehat{A}}\right]}{\Tr e^{-\widehat{A}}}\bigg)\frac{e^{-\widehat{A}}}{\Tr e^{-\widehat{A}}}\nonumber\\
& \simeq \bigg(1+\int_0^1~d\tau~e^{-\tau\widehat{A}}\widehat{B}e^{\tau\widehat{A}}\bigg) \left(1-\int_0^1~d\tau~\langle e^{-\tau\widehat{A}}\widehat{B} e^{\tau\widehat{A}}\rangle_l\right)\widehat{\rho}_l\nonumber\\
& \simeq \bigg[1+\int_0^1 d\tau \bigg\{e^{-\tau\widehat{A}}\widehat{B}e^{\tau\widehat{A}}-\langle e^{-\tau\widehat{A}}\widehat{B}e^{\tau\widehat{A}}\rangle_l\bigg\}\bigg]\widehat{\rho}_l.
\end{align}
here $\langle \widehat{\mathcal{O}}\rangle_l=\Tr\left(\widehat{\rho}_l\widehat{\mathcal{O}}\right)$ 
is the statistical average of the operator $\widehat{\mathcal{O}}$ for the local equilibrium operator.
\section{Derivation of Eq.~\eqref{equ53ver1}}
\label{appenE}
We start with the expression of the operator $\widehat{C}(\vec{x}^{},t^{})$,
\begin{align}
\widehat{C}(\vec{x}^{\prime},t^{\prime}) & =\widehat{T}^{\mu\nu}(\vec{x}^{\prime},t^{\prime})\partial_{\mu}\beta_{\nu}(\vec{x}^{\prime},t^{\prime})-\widehat{J}^{\mu}(\vec{x}^{\prime},t^{\prime})\partial_{\mu}\alpha(\vec{x}^{\prime},t^{\prime})-\widehat{S}^{\mu\rho\sigma}(\vec{x}^{\prime},t^{\prime})\partial_{\mu}\Omega_{\rho\sigma}(\vec{x}^{\prime},t^{\prime})\nonumber\\
& = \widehat{\varepsilon}D\beta-\widehat{P}\beta\theta-\beta\widehat{h}^{\mu}\left(\beta\nabla_{\mu}T-Du_{\mu}\right)+\beta\widehat{\pi}^{\mu\nu}\sigma_{\mu\nu}+\beta\widehat{\Pi}\theta-\widehat{n}D\alpha-\widehat{J}^{\mu}_{(1)}\nabla_{\mu}\alpha-\widehat{S}^{\rho\sigma}D\Omega_{\rho\sigma}-\widehat{S}^{\mu\rho\sigma}_{(1)}\nabla_{\mu}\Omega_{\rho\sigma}.
\label{appenE1ver1}
\end{align}
The spin hydrodynamic equations are, 
\begin{align}
& D\varepsilon+(\varepsilon+P)\theta = -u_{\nu}\partial_{\mu}T^{\mu\nu}_{(1)}\\
& (\varepsilon+P)Du^{\alpha}=\nabla^{\alpha}P-\Delta^{\alpha}_{~\nu}\partial_{\mu}T^{\mu\nu}_{(1)}\\
& Dn+n\theta=-\partial_{\mu}J^{\mu}_{(1)}\\
& D S^{\alpha\beta}+S^{\alpha\beta}\theta=-\partial_{\mu}S^{\mu\alpha\beta}_{(1)}.
\end{align}
Moreover, the thermodynamic relation, 
\begin{align}
dP & =sdT+nd\mu+S^{\alpha\beta}d\omega_{\alpha\beta}\nonumber\\
& =-\frac{1}{\beta}(\varepsilon+P)d\beta+\frac{n}{\beta}d\alpha+\frac{S^{\alpha\beta}}{\beta}d\Omega_{\alpha\beta}.
\label{equE6ver1}
\end{align}
To obtain the above equation, we have used the following relations, 
\begin{align}
dT=-\frac{1}{\beta^2}d\beta;~~d\mu=\frac{1}{\beta}d\alpha-\frac{\alpha}{\beta^2}d\beta;~~d\omega_{\alpha\beta}=\frac{1}{\beta}
 d\Omega_{\alpha\beta}-\frac{1}{\beta^2}\Omega_{\alpha\beta}d\beta.
\label{equE7ver1}
\end{align}
Using Eqs.~\eqref{equE6ver1}-\eqref{equE7ver1} it can be shown that, 
\begin{align}
D\beta=\beta\theta\gamma;~~D\alpha=\beta\theta\gamma^{\prime};~~D\Omega^{\alpha\beta}=\beta\theta\gamma^{\alpha\beta},
\label{appenE8ver1}
\end{align}
here, 
\begin{align}
& \gamma=\bigg[\left(\frac{\partial\varepsilon}{\partial\beta}\right)^{-1}_{n,S_{\alpha\beta}}\left(\frac{\partial P}{\partial\beta}\right)_{\alpha,\Omega_{\alpha\beta}}-\left(\frac{\partial n}{\partial\beta}\right)^{-1}_{\varepsilon,S_{\alpha\beta}}\left(\frac{\partial P}{\partial\alpha}\right)_{\beta,\Omega_{\alpha\beta}}-\left(\frac{\partial S_{\alpha\beta}}{\partial\beta}\right)^{-1}_{n,\varepsilon}\left(\frac{\partial P}{\partial\Omega^{\alpha\beta}}\right)_{\alpha,\beta}\bigg]
\label{equE9new}
\end{align}
\begin{align}
& \gamma^{\prime}= \bigg[\left(\frac{\partial\varepsilon}{\partial\alpha}\right)^{-1}_{n,S_{\alpha\beta}}\left(\frac{\partial P}{\partial\beta}\right)_{\alpha,\Omega_{\alpha\beta}}-\left(\frac{\partial n}{\partial\alpha}\right)^{-1}_{\varepsilon,S_{\alpha\beta}}\left(\frac{\partial P}{\partial\alpha}\right)_{\beta,\Omega_{\alpha\beta}}-\left(\frac{\partial S_{\alpha\beta}}{\partial\alpha}\right)^{-1}_{n,\varepsilon}\left(\frac{\partial P}{\partial\Omega^{\alpha\beta}}\right)_{\alpha,\beta}\bigg]
\label{equE10new}
\end{align}
\begin{align}
& \gamma^{\alpha\beta}=\bigg[\left(\frac{\partial\varepsilon}{\partial\Omega^{\alpha\beta}}\right)^{-1}_{n,S_{\alpha\beta}}\left(\frac{\partial P}{\partial\beta}\right)_{\alpha,\Omega_{\alpha\beta}}-\left(\frac{\partial n}{\partial\Omega^{\alpha\beta}}\right)^{-1}_{\varepsilon,S_{\alpha\beta}}\left(\frac{\partial P}{\partial\alpha}\right)_{\beta,\Omega_{\alpha\beta}}-\left(\frac{\partial S_{\alpha^{\prime}\beta^{\prime}}}{\partial\Omega^{\alpha\beta}}\right)^{-1}_{n,\varepsilon}\left(\frac{\partial P}{\partial\Omega^{\alpha^{\prime}\beta^{\prime}}}\right)_{\alpha,\beta}\bigg].
\label{equE11new}
\end{align}
Now let us look into the term $-\beta \widehat{h}^{\mu}\left(\beta\nabla_{\mu}T-Du_{\mu}\right)$ that appears in Eq.~\eqref{appenE1ver1}, 
\begin{align}
-\beta \widehat{h}^{\mu}\left(\beta\nabla_{\mu}T-Du_{\mu}\right) & = -\beta \widehat{h}^{\mu}\left(\beta\nabla_{\mu}T-\frac{\nabla_{\mu}P}{\varepsilon+P}+\mathcal{O}(\partial^2)\right)\nonumber\\
& \simeq -\beta \widehat{h}^{\mu}\left(\beta\nabla_{\mu}T-\frac{s\nabla_{\mu}T}{\varepsilon+P}-\frac{n\nabla_{\mu}\mu}{\varepsilon+P}-\frac{S^{\alpha\beta}\nabla_{\mu}\omega_{\alpha\beta}}{\varepsilon+P}\right)\nonumber\\
& \simeq -\beta \widehat{h}^{\mu}\left(\beta\frac{\mu n+S^{\alpha\beta}\omega_{\alpha\beta}}{\varepsilon+P}\nabla_{\mu}T-\frac{n}{\varepsilon+P}\nabla_{\mu}\mu-\frac{S^{\alpha\beta}}{\varepsilon+P}\nabla_{\mu}\omega_{\alpha\beta}\right)\nonumber\\
& \simeq \widehat{h}^{\mu} \left(\frac{n}{\varepsilon+P}\nabla_{\mu}\alpha+\frac{S^{\alpha\beta}}{\varepsilon+P}\nabla_{\mu}\Omega_{\alpha\beta}\right).
\label{equE12ver1}
\end{align}
To obtain the final expression, we have used thermodynamic relations and Eq.~\eqref{equE7ver1}. Moreover, we have neglected higher-order derivative terms. Using Eqs.~\eqref{appenE8ver1}, and \eqref{equE12ver1} back in Eq.~\eqref{appenE1ver1} we find, 
\begin{align}
\widehat{C}(\vec{x}^{\prime},t^{\prime}) & = -\left(\widehat{P}-\widehat{\Pi}-\widehat{\varepsilon}\gamma+\widehat{n}\gamma^{\prime}+\widehat{S}^{\alpha\beta}\gamma_{\alpha\beta}\right)\beta\theta
-\left(\widehat{J}^{\mu}_{(1)}-\frac{n}{\varepsilon+P}\widehat{h}^{\mu}\right)\nabla_{\mu}\alpha+\widehat{h}^{\mu}\frac{S^{\alpha\beta}}{\varepsilon+P}\nabla_{\mu}(\beta\omega_{\alpha\beta})+\beta \widehat{\pi}^{\mu\nu}\sigma_{\mu\nu}\nonumber\\
& -2\widehat{\Phi} u^{\alpha}\nabla^{\beta}(\beta\omega_{\alpha\beta})-2\widehat{\tau}^{\mu\beta}_{(s)}u^{\alpha}\Delta^{\gamma\rho}_{\mu\beta}\nabla_{\gamma}(\beta\omega_{\alpha\rho})-2\widehat{\tau}^{\mu\beta}_{(a)}u^{\alpha}\Delta^{[\gamma\rho]}_{[\mu\beta]}\nabla_{\gamma}(\beta\omega_{\alpha\rho}) -\widehat{\Theta}_{\mu\alpha\beta}\Delta^{\alpha\delta}\Delta^{\beta\rho}\Delta^{\mu\gamma}\nabla_{\gamma}(\beta\omega_{\delta\rho})\nonumber\\
& =  -\widehat{P}^{\star}\beta\theta
-\widehat{\mathcal{J}}^{\mu}\nabla_{\mu}\alpha+\widehat{h}^{\mu}\frac{S^{\alpha\beta}}{\varepsilon+P}\nabla_{\mu}(\beta\omega_{\alpha\beta})+\beta \widehat{\pi}^{\mu\nu}\sigma_{\mu\nu}\nonumber\\
& -2\widehat{\Phi} u^{\alpha}\nabla^{\beta}(\beta\omega_{\alpha\beta})-2\widehat{\tau}^{\mu\beta}_{(s)}u^{\alpha}\Delta^{\gamma\rho}_{\mu\beta}\nabla_{\gamma}(\beta\omega_{\alpha\rho})-2\widehat{\tau}^{\mu\beta}_{(a)}u^{\alpha}\Delta^{[\gamma\rho]}_{[\mu\beta]}\nabla_{\gamma}(\beta\omega_{\alpha\rho}) -\widehat{\Theta}_{\mu\alpha\beta}\Delta^{\alpha\delta}\Delta^{\beta\rho}\Delta^{\mu\gamma}\nabla_{\gamma}(\beta\omega_{\delta\rho}).
\label{appenE13ver1}\\
&= -\widehat{P}^{\star}\beta\theta
-\widehat{\mathcal{J}}^{\mu}\nabla_{\mu}\alpha+\widehat{h}^{\mu}\frac{S^{\alpha\beta}}{\varepsilon+P}\nabla_{\mu}(\beta\omega_{\alpha\beta})+\beta \widehat{\pi}^{\mu\nu}\sigma_{\mu\nu}-\widehat{S}^{\mu\alpha\beta}_{(1)}\nabla_{\mu}(\beta\omega_{\alpha\beta}).\label{Cf}
\end{align}
In the above equation,
\begin{align}
& \widehat{P}^{\star}=\left(\widehat{P}-\widehat{\Pi}-\widehat{\varepsilon}\gamma+\widehat{n}\gamma^{\prime}+\widehat{S}^{\alpha\beta}\gamma_{\alpha\beta}\right),\\
& \widehat{\mathcal{J}}^{\mu} = \widehat{J}^{\mu}_{(1)}-\frac{n}{\varepsilon+P}\widehat{h}^{\mu}.
\end{align}

\bibliography{ref.bib}{}
\bibliographystyle{utphys}
\end{document}